\documentclass[a4paper,11pt]{article}
\pdfoutput=1
\usepackage[utf8x]{inputenc}
\usepackage{jheppub}
\usepackage{epsf}
\usepackage{graphicx}	
\usepackage{dcolumn}	
\usepackage{bm}			
\usepackage{amssymb, amsmath}
\usepackage{wasysym}
\usepackage{lipsum, color}

\usepackage{bbm}		

\usepackage{slashed}
\usepackage{hyperref}
\hypersetup{colorlinks, citecolor=bluscuro, linkcolor=black, urlcolor=bluscuro}
\definecolor{rossos}{cmyk}{0,1,1,0.55}
\definecolor{bluscuro}{rgb}{0.15, 0.2, .85}
\definecolor{bluchiaro}{cmyk}{1,.3,0.,0.1}

\let\oldquote\quote
\renewcommand\quote{\scriptsize\oldquote}
\let\oldquotation\quotation
\renewcommand\quotation{\scriptsize\oldquotation}

\newcommand{\gsim}{\lower.7ex\hbox{$\;\stackrel{\textstyle>}{\sim}\;$}}
\newcommand{\lsim}{\lower.7ex\hbox{$\;\stackrel{\textstyle<}{\sim}\;$}}

\def\beq{\begin{equation}}
\def\eeq{\end{equation}}
\def\bea{\begin{eqnarray}}
\def\eea{\end{eqnarray}}
\def\bmat{\begin{pmatrix}}
	\def\emat{\end{pmatrix}}
\def\bei{\begin{itemize}}
	\def\eei{\end{itemize}}

\def\arg{\, {\rm arg}}

\usepackage[usenames,dvipsnames,svgnames]{xcolor}
\usepackage[normalem]{ulem}

\makeatletter
\def\section{\@startsection {section}{1}{\z@}{-3.5ex plus -1ex minus
		-.2ex}{2.3ex plus .2ex}{\large\bf}}
\def\subsection{\@startsection{subsection}{2}{\z@}{-3.25ex plus -1ex
		minus -.2ex}{1.5ex plus .2ex}{\normalsize\bf}}
\makeatother
\makeatletter

\@addtoreset{equation}{section}

\makeatother

\usepackage{graphicx}  
\usepackage{dcolumn}   
\usepackage{bm,relsize}        
\usepackage{amssymb, amsmath}
\usepackage{wasysym}
\usepackage{slashed}
  \usepackage{bbold} 
\usepackage[usenames,dvipsnames,svgnames]{xcolor}

\usepackage{mathtools}
\usepackage{comment}
\usepackage{cancel}
\usepackage{ulem}

\usepackage{feyn}

\def\beq{\begin{equation}}
\def\eeq{\end{equation}}

\providecommand{\tabularnewline}{\\}
\begin{document}

\global\long\def\com#1#2{\underset{{\scriptstyle #2}}{\underbrace{#1}}}

\global\long\def\comtop#1#2{\overset{{\scriptstyle #2}}{\overbrace{#1}}}

\global\long\def\ket#1{\left|#1\right\rangle }

\global\long\def\bra#1{\left\langle #1\right|}

\global\long\def\braket#1#2{\left\langle #1|#2\right\rangle }

\global\long\def\op#1#2{\left|#1\right\rangle \left\langle #2\right|}

\global\long\def\opk#1#2#3{\left\langle #1|#2|#3\right\rangle }

\global\long\def\L{\mathcal{L}}

\title{High quality Nelson-Barr solution to the strong CP problem with $\theta=\pi$}

\author[a]{Gilad Perez,}
\author[a]{Aviv Shalit}
\affiliation[a]{Weizmann Institute of Science, Rehovot 76100, Israel}
\date{\today}

\abstract{We discuss composite UV completions of the Nelson-Barr(NB) solution to the strong CP problem. In our construction, the CP symmetry is broken spontaneously by the dynamics of a hidden QCD at $\theta=\pi$. We focus on the minimal implementation of the NB construction where the visible sector contains one extra pair of vector-like up/down quarks. 
We show that the minimal NB theory suffers from a quality problem, and discuss how composite UV completions may resolve it. We present a simple calculable scheme, free of a quality problem, where dynamical  CP violation in the hidden sector is mediated through a scalar portal to the visible sector, which successfully realizes the NB construction.}

\maketitle

\section{Introduction}
The strong CP problem in the Standard Model (SM) arises from the vast hierarchy between its two allowed CP violating phases: the strong CP phase, originating from the combination of the bare $\theta_{\text{QCD}}$ angle and the argument of the determinant of the Yukawa mass matrices, and the CKM phase, arising as a consequence of the misalignment between the Yukawas of the up and the down sector, $Y_u$ and $Y_d$ respectively. In a basis independent parametrization \cite{Jarlskog:1985ht} we can write   
\begin{align}
&\bar \theta_{\text{QCD}}=\theta_{\text{QCD}}+\text{arg}\det(Y_uY_d)\lesssim 10^{-10}\, ,\label{eq:strongCP}\\
&\theta_{\text{weak}}=\text{arg}\text{ det}[Y_u Y_u^\dagger,Y_dY_d^\dagger] \simeq \mathcal{O}(1)\, ,
\end{align}
where the upper bound on $\bar{\theta}$ comes from neutron and Hg dipole moments measurements \cite{Baker:2006ts,2016PhRvL.116p1601G,Afach:2015sja} and $\theta_{\text{weak}}\simeq 1$ is the CKM, CP-violating phase, which controls the CP-violating processes in the meson systems \cite{Hocker:2006xb}. 

Generally speaking, one can identify two classes of solutions to the strong CP problem: the Peccei-Quinn (PQ) solutions \cite{Peccei:1977aa} and the solutions involving spontaneous breaking of CP \cite{Nelson:1983zb,Barr:1984qx,Barr:1984fh,Bento:1991ez,Vecchi:2014hpa,Dine:2015jga,Schwichtenberg:2018aqc}.\footnote{We focus here on constructions where CP is spontaneously broken, alternative models involving spontaneous parity violation were discussed in \cite{Babu:1989rb,Barr:1991qx,Kuchimanchi:2010xs}.} The PQ solution relies on a global symmetry $U(1)_{\text{PQ}}$ which is spontaneously broken at a scale $f_a$ and anomalous under QCD. Below $f_a$ the strong CP phase is controlled by the dynamics of the the $U(1)_{\text{PQ}}$ Goldstone (the axion) which below the QCD confinement scale gets a potential stabilizing the strong CP phase to zero \cite{Kim-Axion}. This solution gives a dark matter candidate for free and a low-energy experimental target to hunt for \cite{Kim-Axion}. At present, astrophysical bounds give $f_a\gtrsim 10^8\text{ GeV}$ and the planned experimental effort will allow us to test quite extensively the allowed parameter space in the next decade \cite{Graham:2015ouw}. 

On the theory side, the axion solution has a certain fragility which stems from the fact that the $U(1)_{\text{PQ}}$ is required to be spontaneously broken at high scale $(f_a\gtrsim 10^8\text{ GeV})$ and explicitly broken only by the QCD anomaly. Any other operator breaking $U(1)_{\text{PQ}}$ explicitly tends to push the axion vacuum expectation value (VEV) away from the origin. This problem is exacerbated at high $f_a$ where stabilizing the axion close to the origin requires an incredibly precise accident of the UV theory, either suppressing the Wilson coefficient of the $U(1)_{\text{PQ}}$-breaking operators \cite{Kallosh:1995hi} or forbidding all $U(1)_{\text{PQ}}$-breaking operators up to high dimension \cite{Randall:1992ut,Cheng:2001ys,Hill:2002kq,Choi:2003wr,Flacke:2006ad,Redi:2016esr}. Thus, we say that axion solution suffers from a \emph{quality problem} \cite{Georgi:1981pu,Lazarides:1985bj,Kamionkowski:1992mf,Holman:1992us,Barr:1992qq} which correspond to the difficulty of making the Peccei-Quinn symmetry an accident of the low energy dynamics. This theoretical issue might be a red herring, or perhaps a reason to look at alternative frameworks.

The second class of solutions makes use of the fact that within the SM the CKM phase renormalizes the strong CP phase only at seven loops \cite{Ellis:1978hq,Khriplovich:1993pf} and finite threshold corrections from quark masses are also very suppressed \cite{Khriplovich:1985jr}, opening up the possibility of having purely UV solution to the strong CP problem\footnote{The argument given in \cite{Khriplovich:1993pf} to understand the seven loops suppression of the RGE contribution goes as follows: i) $J$ is the first flavor singlet which can contribute to the log-divergent part of $\bar \theta_{\text{QCD}}$ and $J\sim y^{12}$ so we need to close the 6 higgs loops ii) $u\to d$ implies $J\to -J$ while $\bar \theta_{\text{QCD}}$ is left unchanged. Therefore we need to break this symmetry with an extra gauge loop which distinguish up and down quarks in order to get the first non zero contribution to $\bar \theta_{\text{QCD}}$.}. In this type of solutions CP is assumed to be a symmetry of the UV theory, possibly motivated by string theory embeddings \cite{Dine:1992ya,Banks:2010zn}. CP is then spontaneously broken in a hidden sector and this breaking is communicated to the SM through a flavorful portal. The main challenge is to generate a order one CKM phase while preserving the suppression of the $\bar \theta_{\text{QCD}}$ parameter. The two main realizations of this idea are differentiated by how the $\bar \theta_{\text{QCD}}$ parameter is protected. One possibility is to make the Yukawas hermitian either by introducing supersymmetry \cite{Hiller_2002} or extra dimensions \cite{Harnik_2005}. The other possibility, originally put forward by Nelson \cite{Nelson:1983zb} and generalized by Barr \cite{Barr:1984qx,Barr:1984fh}, introduces a set of (discrete) symmetries which force a special texture of the fermions mass matrix forbidding the strong CP-phase while allowing the CKM one. 

\begin{figure}[h!]
	\centering
	\includegraphics[width=1\textwidth]{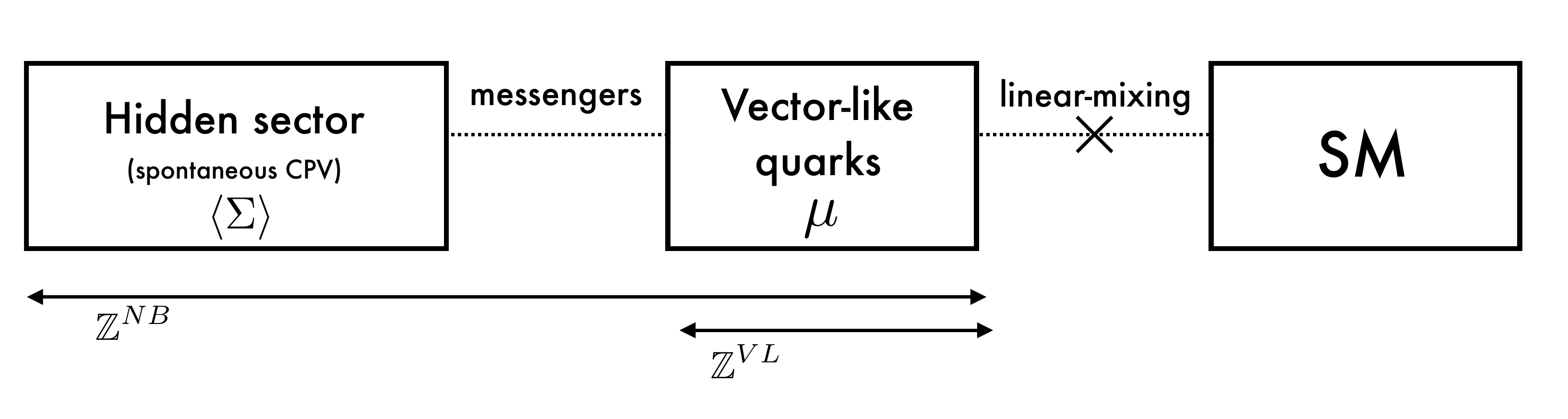}
	\caption{Cartoon of the structure of Nelson-Barr solutions. The CP-symmetry is broken in a hidden sector that communicates to the vector-like quarks only. The latter are linearly mixed with the SM $SU(2)_L$ singlet quarks (up and/or down) such that a the CKM phase is generated at tree level while $\bar{\theta}$ is suppressed. A set of (discrete) symmetries ensures the suppression of $\bar \theta_{\text{QCD}}$. These can act on both the hidden sector and the vector-like quarks ($\mathbb{Z}^{NB}$ in the cartoon) or on the vector like fermions only ($\mathbb{Z}^{VL}$ in the cartoon).}
	\label{fig:NBstructure}
\end{figure}

The general structure of Nelson-Barr (NB) solutions is depicted in Fig.~\ref{fig:NBstructure}. 
A hidden sector is assumed to break CP spontaneously through the VEV of a CP-odd operator, $\langle\Sigma\rangle$. The visible sector is instead extended by the addition of new fermions that linearly mixes with the SM $SU(2)_L$ singlet quarks (up and/or down) and have a vector-like mass $\mu$ (see also \cite{Vecchi:2014hpa,Egana-Ugrinovic:2018znw,Egana-Ugrinovic:2019dqu}). CP violation is then communicated to the visible sector (vector-like fermions plus SM fermions) through a portal and the CKM phase is generated at the tree level, while a set of symmetries suppresses the contributions to the strong CP phase. These symmetries might act on the vector-like fermions and the hidden sector ($\mathbb{Z}^{\rm NB}$ in the figure) or on the vector-like fermions only ($\mathbb{Z}^{\rm VL}$ in the figure).
The simplest implementation of the NB setup, the Branco-Bento-Parada (BBP) model \cite{Bento:1991ez} will be reviewed in Sec.~\ref{sec:BBP-Intro}. In this minimal model, CP is broken by the VEV of a single fundamental scalar and it is mediated to the visible sector through a direct Yukawa couplings which are triplet under the $U(3)_{u,d}$ flavor group. A non-anomalous $\mathbb{Z}_2^{\text{NB}}$-symmetry acting on both the scalar and the vector-like fermions forbids the strong CP phase at tree level, but controlling loop corrections requires these coupings to be smaller than $10^{-3}$. In addition, generating an $\mathcal{O}(1)$ CKM phase requires the VEV of the fundamental to satisfy: $\kappa\langle\Sigma\rangle\gtrsim \mu$, where $\kappa$ is the scale of the new portal Yukawa couplings and $\mu$ is bounded from below by flavor bounds and collider constraints. Such a high scale VEV for the scalar, makes the theory sensitive to higher-dimensional operators which are allowed by the non-anomalous $\mathbb{Z}_2^{\text{NB}}$ symmetry and can be suppressed only by tuning the fundamental scalar VEV to be small with respect to the cut-off of the theory $\Lambda$. As first noticed in~\cite{Dine:2015jga}, even if one pushes the cut-off to be as high as $M_{\text{Pl}}$ the tuning required to have have a fundamental scalar breaking CP at the scale $f_{\text{CP}}\ll M_{\text{Pl}}$ is much worst than the original hierarchy between $\bar \theta_{\text{QCD}}$ and $\theta_{\text{weak}}$. Here we find that in the simple BBP model, Planck suppressed operators alone would completely destroy the NB texture unless their Wilson coefficients are accidentally small. A UV completion is then necessary to address the NB quality problem and assess the viability of the NB solution.  

We explore composite UV completions of the NB mechanism where CP is broken by a hidden $SU(N)$ Yang Mills (YM) theory at $\theta'=\pi$ with $N_f$ flavors. Depending on the hierarchy between the confinement scale and the quark masses we explore the possibility that CP is broken by the chiral condensate or by the glueball condensate. In the former case the breaking of CP its evident from the chiral lagrangian at large $N$~\cite{Dashen:1970et,Witten:1980sp,DiVecchia:2017xpu} while for the latter case recent results ensures that CP is broken even at finite $N$~\cite{Gaiotto:2017yup,Gaiotto:2017tne}. The CP-breaking hidden sector is connected to the visible sector through a scalar portal only, while the visible sector has the typical BBP structure ensured by a $\mathbb{Z}_2^{\text{NB}}$ symmetry which acts non-trivially both on the hidden and the visible fermions. 
The discrete $\mathbb{Z}_2^{\text{NB}}$, which controls the structure of the visible sector, is only spontaneously broken in the original BBP model~\cite{Bento:1991ez}, while here an explicit soft breaking will be added.\footnote{This symmetry can be extended to higher rank groups~\cite{Dine:2015jga} and continuous groups such as flavor symmetry as shown in~\cite{Vecchi:2014aa}. In the interest of minimality, we will restrict our attention to a simple $\mathbb{Z}_2^{\text{NB}}$-symmetry acting on a single CP-breaking spurion. One could envisage more complicated constructions with multiple spurions where CP is broken together with flavor symmetries of the standard model \cite{Vecchi:2014hpa,Egana-Ugrinovic:2018znw}. In these extended scenarios one could hope to address the flavor puzzle and the strong CP problem together \cite{Davidson:2007si,Grinstein:2010ve,Nardi:2011st,Espinosa:2012uu,Fong:2013dnk}.} Crucially, dangerous operators spoiling the NB texture are further suppressed by a $Z_2^{\text{VL}}$ which is only broken by the vector-like fermions mass scale $\mu$.

This paper is organized as follows. In Section~\ref{sec:BBP-Intro} we review the BBP model and the constraints on its parameters space: i) from making the CKM phase of order one (Sec.~\ref{sec:CKMphase}) ii) from suppressing the strong CP phase below the present experimental bound (Sec.~\ref{sec:radcorr}) iii) from having the new vector-like fermions above the present flavor bounds (Sec.~\ref{sec:flavorconst}). All these constraints together lead to the Nelson-Barr quality problem illustrated in Sec.~\ref{sec:NBquality}. In Sec.~\ref{sec:UVcompletions} we discuss composite UV completions of NB with a hidden $\theta'=\pi$ parameter. We first present a hidden chiral model as an extension to the minimal BBP model. However this model has a small allowed parameter space and thus considered accidental. later, we show a NB model based on a pure YM theory at $\theta'=\pi$. This type of models exhibit a large range of allowed parameter space and thus theoretically favorable. In Sec.~\ref{sec:conclusions} we conclude our results and address the solution to the quality problem withing these composite NB models. In Appendix~\ref{app:spontaneousCP} we collect theory results supporting the fact that CP is spontaneously broken in pure YM at $\theta'=\pi$.

\section{The simplest Nelson-Barr model and its challenges}\label{sec:BBP-Intro}
Since CP is an exact symmetry in Nelson Barr models, there exists a basis in which all coupling constants are real. In what follows, we will work in this particular basis only to make our analysis manifestly CP conserving. Here we describe a specific NB model first introduced by Bento-Branco-Parada (BBP)~\cite{Bento:1991ez}. This model has the virtue of introducing a minimal set of fields beyond the SM ones. A new complex scalar $\Sigma$ singlet under the SM gauge group, has a potential such that its VEV breaks CP spontaneously. Moreover a pair of vector like fermions $(U,\tilde{U})/(D,\tilde{D})$ in the same representation of the $SU(2)_L$ singlet up/down quarks are introduced.\footnote{In what follows we present the BBP model extending the up quark sector. The same construction would hold in the down sector in exactly the same way just replacing $u\leftrightarrow d$ and $U\leftrightarrow D$. The only difference will appear in the flavor constraints the we will discuss in Sec.~\ref{sec:flavorconst}.}

All the BSM field are odd under a $\mathbb{Z}_2^{\text{NB}}$-symmetry while the SM ones are even 
\begin{equation}
 \mathbb{Z}_2^{\text{NB}}: U\rightarrow -U\, , \tilde{U}\rightarrow -\tilde{U}\, , \Sigma\to -\Sigma\, , \label{eq:Z2NBBBP}
\end{equation}
such that the only renormalizable Yukawa couplings in the up sector are 
\begin{equation}
\L^{u}= Y_{ij}^{u}Q_{i\alpha}\epsilon^{\alpha\beta}H_{\beta}\tilde{u}_{j} +\mu U\tilde{U}+\left(\kappa_{i}^u\Sigma+\tilde{\kappa}_{i}^u\Sigma^{*}\right)U\tilde{u}_{i} + \text{h.c.}\, ,\label{eq:NB_Lagrangian-1NB}
\end{equation}
where $Y_{ij}^{u}$ are the $3\times3$ Yukawa couplings
of the SM up-quark sector ($i,j=1,2,3$), $\mu$ is the mass parameter
of the new heavy vector like fermions and $\kappa_{i}^u$ and $\tilde{\kappa}_{i}^u$,
for $i=1,2,3$, are the NB Yukawa couplings, which couple all three
generations of the $SU(2)_L$ singlet up quarks $\tilde{u}_i$ with the new
scalar(s) $\Sigma$ through the heavy fermion $U$. 

As mentioned, we assume that scalar(s) potential is such that the VEV of $\Sigma$ has a $\mathcal{O}\!\left(1\right)$ phase which breaks CP spontaneously. We can then write
\[
\left\langle \Sigma\right\rangle =\frac{f_{\text{CP}}}{\sqrt{2}}e^{i\eta}\,.
\]
where $f_{\text{CP}}$ controls the scale of the $\Sigma$-fluctuations, while $\eta\sim\mathcal{O}(1)$ phase. The tree-level mass matrix in the up sector is
\begin{equation}
\left(\begin{array}{cccc}
U & Q_{i} \end{array}\right) M_{4\times4}^{u}\left(\begin{array}{c}
\tilde{U}\\
\tilde{u}_{j}
\end{array}\right)=\left(\begin{array}{cccc}
U & Q_{i} \end{array}\right)\left(\begin{array}{cc}
\mu & B_{j}^u\\
0 & vY^{u}_{ij}
\end{array}\right)\left(\begin{array}{c}
\tilde{U}\\
\tilde{u}_{j}
\end{array}\right) \, ,\label{eq:mass_D_tree_level-NB}
\end{equation}
where we defined
\begin{equation}
B_{i}^u\equiv\frac{f_{\text{CP}}}{\sqrt{2}}\left(\kappa_{i}^ue^{i\eta}+\tilde{\kappa}_{i}^ue^{-i\eta}\right)\,.\label{eq:defBu}
\end{equation}
At tree level, $\arg\left(\det M_{4\times4}^{u}\right)=0$ is manifest in this model. Since the bare $\theta_{\text{QCD}}$ parameter also vanishes because of the assumption of CP invariance in the UV, the physical $\bar{\theta}_{\text{QCD}}$ parameter is also vanishing at tree level:
\begin{equation}
\bar{\theta}_{\text{QCD}}=\theta_{\text{0}}+\arg\left(\det M_{4\times4}^{u}\right)=0\,.\label{eq:NBsol_tree_level1}
\end{equation}
The $\mathbb{Z}_{2}^{\text{NB}}$ symmetry in Eq.~\eqref{eq:Z2NBBBP} guarantees the NB mass texture of Eq.~\eqref{eq:mass_D_tree_level-NB} at tree level, which is crucial to achieve the suppression of $\bar{\theta}_{\text{QCD}}$.

\subsection{The CKM phase}\label{sec:CKMphase}
The spontaneous breaking of CP must generate an ${\cal O}\!\left(1\right)$
CKM phase. In order to estimate the CKM phase in the model, we first
integrate out the heavy degrees of freedom, above the weak scale. One can compute the heavy and light mass eigenvalues by diagonalizing
the mass matrix square
\begin{equation}
M_{4\times4}^{u}M_{4\times4}^{u\dagger} =\left(\begin{array}{cc}
\mu^{2}+\left|B_{i}^u\right|^{2} & vB_{k}^uY_{jk}^{u}\\
v Y_{ik}^{u}B_{k}^{u\ast} & v^2Y_{ik}^{u}Y_{jk}^{u}
\end{array}\right)\, ,\label{eq:NBup}
\end{equation}
where the summation over the repeated indices is omitted. In the above expression we use $m_{ij}^u=m_{ij}^{u\ast}$ in our favourite basis, where the only source of CP-violation is in $B_i$. 
In the limit $\mu^{2}+\left|B_{k}^u\right|^{2}\gg\vert m_{ik}^{u}m_{jk}^{u}\vert\;\text{for all } i,j$,
we can approximately diagonalize $M_{4\times4}^{u}M_{4\times4}^{u\dagger}$
and integrate out the heavy fermionic state. This leads to the low energy effective $3\times3$
mass matrix squared:
\begin{equation}
\left(M_{3\times3}^{u}M_{3\times3}^{u\dagger}\right)_{ij}\approx  v^2\left[ Y_{ik}^{u}Y_{jk}^{u}-\frac{Y_{ik}^{u}B_{k}^{u*}B_{\ell}^u Y_{\ell j}^{u}}{\mu^{2}+\left\vert B^u_k\right\vert^{2}}\right]\,,\label{eq:3times3mass_square-NB}
\end{equation} 
Clearly, any phase in the unitary matrix $V_{L}^{u}$ which diagonalizes the matrix above would lead to a phase in the CKM matrix.

If we assume that the SM quark masses and mixings are controlled by the real Yukawa couplings in the up and the down sector (i.e. $\mu\gtrsim v$), obtaining a large CKM phase constrains the parameters of the NB model:
 \begin{itemize}
\item First, one must require that 
\begin{equation}
\mu^{2}\lsim\left| B^u_k\right|^{2}\, ,\label{eq:cond1}
\end{equation}
such that the second term in Eq.~\eqref{eq:3times3mass_square-NB}
is not negligible compare to the first term. For heavy $\mu$ the physical CP phase in the CKM will be suppressed by $\sim \left| B^u_k\right|^{2}/\mu^2$. 
\item Second, the requirement
that the second term in Eq.~\eqref{eq:3times3mass_square-NB} contains
an order one phase is translated into a relation between the NB Yukawa couplings:
\begin{equation}
\frac{\left\vert\vec{\kappa}^u\times\vec{\tilde{\kappa}}^u\right\vert}{\left\vert\vec{\kappa}^u\right\vert^{2}+\left\vert\vec{\tilde{\kappa}}^u\right\vert^{2}}\sim\mathcal{O}(1)\, ,\label{eq:CPviol}
\end{equation}
which correspond to the requirement of not having the two flavor spurions $\vec{\kappa}^u$ and $\vec{\tilde{\kappa}}^u$ aligned in flavor space, see also~\cite{Davidi:2017aa,Davidi:2018sii} .The physical CP phase in the CKM depends linearly on the outer product of the two flavor spurions in Eq.~\eqref{eq:CPviol}. 
\end{itemize}

\subsection{The strong CP phase}\label{sec:radcorr}
In this part,we present the radiative corrections to the BBP model. Indeed, the $\mathbb{Z}_{2}^{\text{NB}}$ symmetry in Eq.~\eqref{eq:Z2NBBBP} is broken together with CP by the VEV of the scalar $\Sigma$ and as a consequence quantum corrections involving the singlet dynamics will unavoidably spoil the NB texture in Eq.~\eqref{eq:mass_D_tree_level-NB}. In order to keep track of the quantum corrections we can write $M^u_{4\times4}\to M^u_{4\times4}+\Delta M^u_{4\times4}$ such that,   
\begin{equation}
\Delta\bar{\theta}_{\text{QCD}}\simeq \text{Im}\left[\text{Tr}\left((M^u_{4\times4})^{-1}\Delta M^u_{4\times4}\right)\right]\, , \label{eq:radiativetheta}
\end{equation}
and 
\begin{equation}
\Delta M^u_{4\times4}=\left(\begin{array}{cc}
\Delta \mu & \Delta B_j^u\\
v\Delta \kappa^{q}_i& v\Delta Y^u_{ij}
\end{array}\right)\, .
\end{equation}
This encodes the quantum corrections to the $4\times4$ up quark matrix in Eq.~\eqref{eq:NBup}. It is useful to write Eq.~\eqref{eq:radiativetheta} explicitly in components to distinguish the scaling of the different radiative corrections: 
\begin{align}
&\Delta\bar{\theta}_{\text{QCD}}\vert_{\Delta\mu}= \frac{1}{\mu}\text{Im}\left[\Delta \mu\right]\,,\label{eq:deltatheta1}\\
&\Delta\bar{\theta}_{\text{QCD}}\vert_{\Delta \kappa^{q}}=  -\frac{1}{\mu}\text{Im}\left[(BY_u^{-1})_i\Delta \kappa^q_i\right]\,,\label{eq:deltatheta2}\\
&\Delta\bar{\theta}_{\text{QCD}}\vert_{\Delta Y^u}=  \text{Im}\left[Y^{u -1}_{ik}\Delta Y^u_{ki}\right]\,.\label{eq:deltatheta3}
\end{align} 
As expected, because of the NB texture in Eq.~\eqref{eq:mass_D_tree_level-NB} any loop correction to the NB Yukawa $B_i$ would not generate a new contribution to $\theta_{\text{QCD}}$. This phase is indeed responsible for the CKM and would feed into the strong CP one only through SM running.  The three dangerous contribution arise from new phases in the vector-like fermion mass $\Delta\mu$ and/or in the quark up Yukawa $\Delta Y^u$, or new contributions $\Delta\kappa^{q}$ (both real and imaginary) to the ``wrong Yukawa'' $HQ\tilde{U}$ which was set to zero by the $\mathbb{Z}_{2}^{\text{NB}}$ symmetry at tree level. 

\begin{figure}[h!]
		\begin{centering}
		\includegraphics[scale=0.23]{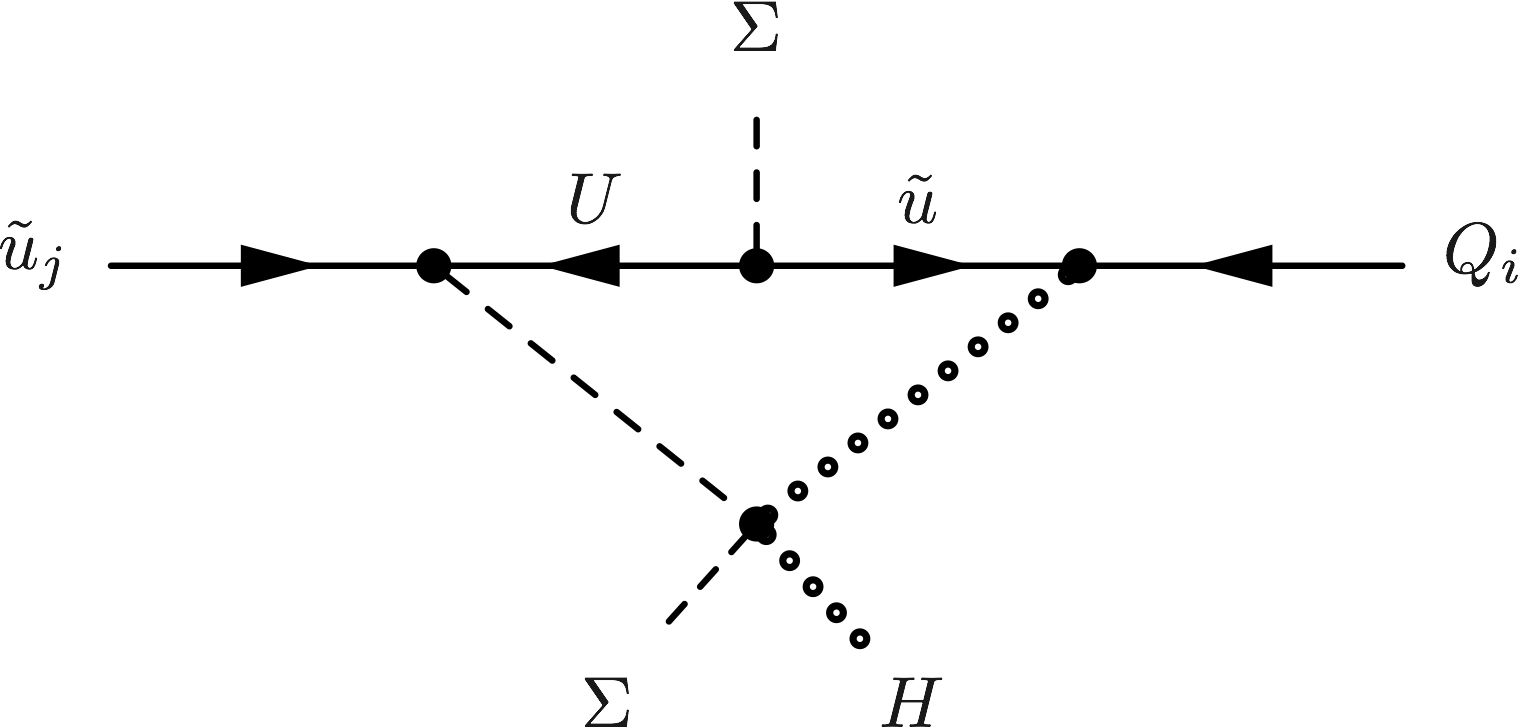}\quad\quad\quad\quad
		\includegraphics[scale=0.23]{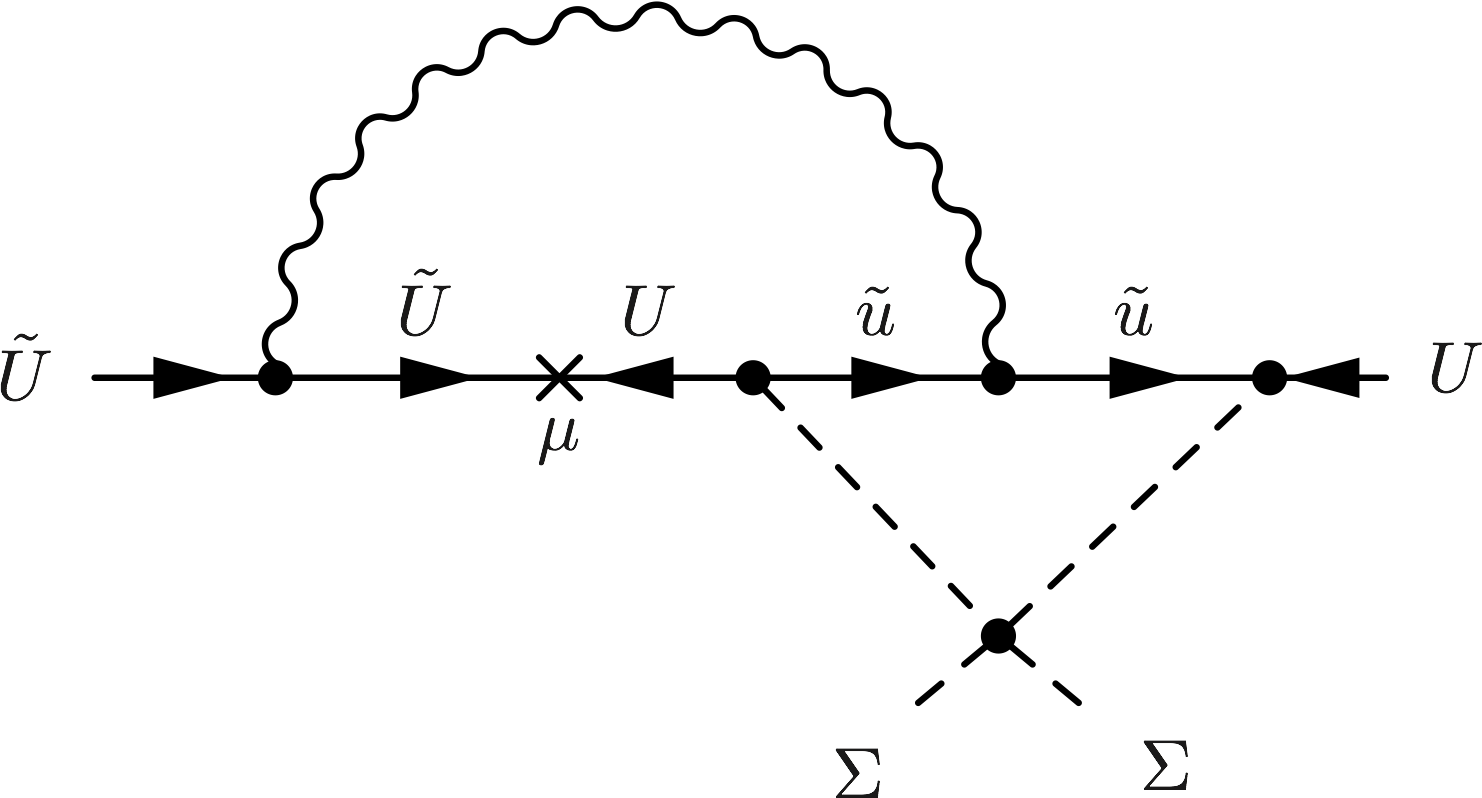}
		\caption{The leading loop corrections to $\bar{\theta}_{\text{QCD}}$ in the BBP model. {\bf Left:} one-loop contributions controlled by the interactions of the CP-violating scalar with the SM Higgs.  {\bf Right:} two-loops ``dead duck'' contributions where only the CP-violating scalar propagates.~\label{fig:Usual_NB_Diagrams}}
		\par \end{centering}
\end{figure}

A subset of the loop corrections generated by the singlet dynamics are depicted in Fig.~\ref{fig:Usual_NB_Diagrams} and depends on its potential and its mixing with the SM Higgs
\begin{equation}
\mathcal{L}\supset\lambda_H \Sigma^{\dagger}\Sigma H^{\dagger}H+ \kappa_H(\Sigma^2+\Sigma^{\dagger2}) H^{\dagger} H+ \kappa_\Sigma (\Sigma^2+\Sigma^{\dagger2})\Sigma^{\dagger}\Sigma+ \gamma_\Sigma (\Sigma^4+\Sigma^{\dagger4})\, .
\end{equation}

As already noticed in \cite{Bento:1991ez}, the contributions due to the Higgs mixing in Fig.~\ref{fig:Usual_NB_Diagrams} left decouple for $f_{\text{CP}}\gg v$ at the price of tuning the Higgs mass to be smaller than the scale of spontaneous CP violation. At 2-loops, one can find ``dead duck'' contributions \cite{Nelson:1983zb,Dine:2015jga} depicted in Fig.~\ref{fig:Usual_NB_Diagrams} right where only the singlet $\Sigma$ propagates. Taking the $\Sigma$ quartics of $\mathcal{O}(1)$, which is necessary to get a $\mathcal{O}(1)$ CP-violating VEV \cite{Haber:2012np} and requiring the CP phase to be less than the present experimental bound one gets an upper bound on the Nelson-Barr Yukawa couplings
\begin{equation}
\vert \vec{k}^u\vert\sim \vert \vec{\tilde{k}}^u\vert\lesssim 10^{-3}\, . \label{eq:quantumcorr}
\end{equation}
In a concrete composite UV completion we will see how the smallness of the NB Yukawa can easily be related to the ratio of the confinement scale over the much higher portal scale. 

\subsection{Flavor  \& collider constraints}\label{sec:flavorconst}
We now move to the flavor constraints on the scale of the vector-like $SU(2)_L$ singlet fermions. The $\mathbb{Z}_{2}^{\text{NB}}$-symmetry which is protecting $\theta_{\text{QCD}}$ at tree level is also suppressing any source of flavor violation at tree level. Integrating out the vector-like fermion at tree level, we obtain the following GIM violating contributions,
\begin{equation}
\mathcal{L}_Z^u\supset\frac{m_Z^2}{g_Z\bar{M}^2}\lambda_i^u\lambda_j^{u\dagger} \left(u_{L i}\gamma_\mu u_{L j}^{\dagger}\right) Z^{\mu}, \qquad \lambda_i^u=\frac{(B^{u\ast}V^{\dagger}Y_u^\dagger)_i}{\bar{M}}\, , \label{eq:spurion}
\end{equation}
where $\bar{M}\simeq \sqrt{\mu^{2}+\left|B_{i}^u\right|^{2}}$ is the mass of the  new heavy quark generation. \\
This is drastically different from generic vector like fermion models \cite{Nir:1999mg,Ishiwata:2015cga}, where the mixing with the left-handed SM quarks is induced at tree level via the ``wrong'' yukawa coupling $HQ\tilde{U}$, leading to $\lambda_i^u\sim\mathcal{O}(1)$. The ``wrong'' yukawa is forbidden by the $\mathbb{Z}_{2}^{NB}$-symmetry and the Flavor Violation (FV) is then proportional to the SM yukawas and the CKM entries.  The suppression of tree level Flavor Changing Neutral Currents (FCNCs) can be also seen by rotating to the mass eigenstates directly using the full $4\times4$ matrix.  Consider  the 4 generation mass matrix in our inteaction basis as written in Eq.~\eqref{eq:NBup}. One can write this $4\times4$ matrix in the mass (diagonal) basis instead:
\begin{equation}
U_L^{\dagger} M_{4\times4}^{u}M_{4\times4}^{u\dagger} U_L= \left.M_{4\times4}^{u}M_{4\times4}^{u\dagger} \right\vert_{\text{diag}}
\end{equation}
where we define 
\begin{equation} \label{eq:U_L_transformation}
U_L=\left(\begin{array}{cc}
T & S_i\\
R_i & V_{ij}
\end{array}\right)\, ,\qquad S_i\simeq - \frac{ vY_u V B^u}{\mu^2+\vert \vec{B}^u\vert^2}  \, .
\end{equation}
In the above equation, the flavor vector $S_i$ controls the mixing between the left-handed up quarks and the heavy vector-like fermions whose mass squared is $\mu^2+\vert \vec{B}\vert^2$. Tree-level FCNC's are controlled by the non-unitarity of the $3\times3$ CKM-like matrix, $V$, which can be written as 
\begin{equation}
V^{\dagger}_{ik}V_{kj}= \delta_{ij}+S_{i}^{\ast}S_{j}\simeq \delta_{ij}+\frac{v^2(B^{u\ast}V^{\dagger}Y_u^\dagger )_i(Y_u V B^u)_j}{(\mu^2+\vert \vec{B}^u\vert^2)^2} \ . 
\end{equation}
\\
The same $\mathbb{Z}_{2}^{NB}$ forbids the coupling of $\Sigma U\tilde{U}$, which induces a four fermi operator among the vector-like quarks via $\Sigma$-exchange. FV four-fermi operators are also generated at 1-loop in the BBP model when the singlet $\Sigma$ propagates in the loop. As we discussed in the previous section, the BBP model requires $\vert\vec\kappa^u\vert\sim\vert\vec{\tilde{\kappa}}^u\vert\lesssim 10^{-3}$ in order to not generate a too large neutron dipole moment so that any effect where the scalar $\Sigma$ propagates into the loop becomes irrelevant and the spurion limit for the singlet captures the dominant FV effects.

The FV structure controlling the flavor bounds from $c\to u$ transitions is then proportional to 
\begin{equation}
\lambda_1^u(\lambda_2^u)^*\simeq V_{13}V^*_{32} y_t^2\simeq 4\cdot 10^{-4}\ ,
\end{equation}
where we took $\mu\sim \vert B^u\vert$ as required by Eq.~\eqref{eq:cond1}.
Recasting the $\Delta F=1$ bounds of \cite{Ishiwata:2015cga} in our framework we get 
$\mu^2+\vert \vec{B}^u\vert^2\gtrsim (50\text{ GeV})^2\ ,$ where the bound arises from the LHCb bound on $\text{BR}(D_0\to\mu^+\mu^-)\lesssim 6\cdot 10^{-9}$ at 90\%~CL~\cite{Aaij:2013cza}. This observable is however affected by short distance uncertainties. A more robust bound can be obtained looking at $\Delta F=2$ operators induced by tree-level Z-exchange which scale as $(\lambda_i^u\lambda_j^{u\dagger})^2v^2/\bar{M}^4$. Rescaling the bound from $D^0-\bar{D}^0$ mixing obtained in \cite{Ishiwata:2015cga} we get 
\begin{equation}
\mu^2+\vert \vec{B}^u\vert^2\gtrsim (200\text{ GeV})^2\, .\label{eq:flavorup}
\end{equation} 
A similar analysis can be performed for the down sector Nelson-Barr model where we have 
\begin{equation}
\mathcal{L}_Z^d\supset\frac{m_Z^2}{g_Z\bar{M}^2}\lambda_i^d\lambda_j^{d\dagger} \left(d_{L i}\gamma_\mu d_{L j}^{\dagger}\right) Z^{\mu}\ ,\qquad \lambda_i^d=\frac{(B^{d\ast} V^{\dagger}Y_d^\dagger)_i}{\bar{M}}\, . \label{eq:spurion2}
\end{equation}
Here, the relevant FV structures are heavily suppressed by $y_b^2$. Both $\Delta F=1$ and $\Delta F=2$ operators are in principle constrained by precise measurements of rare decays of the kaons and the B mesons $\text{BR}(K_L\to\mu^{+}\mu^{-})<(2.8\pm0.6)\cdot10^{-9}$ \cite{estimateBettler} and $\text{BR}(K^{+}\to\pi^{+}\nu\bar{\nu})<(1.7\pm1.1)\cdot10^{-10}$ \cite{Artamonov:2008qb} and neutral mesons oscillations. Because of the $y_b^2$ suppression these bounds end up constraining very mildly the scale of the down vector-like fermions
\begin{equation}
\mu^2+\vert \vec{B}^d\vert^2\gtrsim (20\text{ GeV})^2\,  .
\end{equation} 

Finally we discuss LHC bounds on the Nelson-Barr fermions.  These end up being the dominant constraint for both up and down sector Nelson-Barr fermions. After rotating into the mass basis, we can define a heavy up quark $T$ with charge $2/3$ and mass $m_T$ and a down quark $B$ with charge $-1/3$ and mass $m_B$ for the up and down Nelson-Barr model respectively. The heavy $T$ quarks are produced in pairs in $pp$ collision at the LHC and then decay with branching ratios $\text{BR}(T\to Zt)\simeq\text{BR}(T\to ht)\simeq1/2\,\text{BR}(T\to Wb)$ as long as they are heavier than the EW scale. For $B$ quarks once should replace $t\leftrightarrow b$ but the same branching ratio hierarchy holds. Notice that the $T$ and $B$ quarks in our setup have a reduced width compared to standard vector-like quarks which is controlled by the $\lambda_i^{u,d}$ spurions. As a consequence single production in association with $b$ and $t$-quarks is suppressed compare to the standard vector-like fermions \cite{DeSimone:2012fs}. Nevertheless $T$ and $B$ still decay promptly on detector scale and the standard bounds from pair production applies. CMS and ATLAS performed a combined analysis based on an integrated luminosity of $ 35.9\text{ fb}^{-1}$ and $36.1 \text{ fb}^{-1}$ at $\sqrt{s}=13\text{ TeV}$ respectively \cite{Sirunyan:2018omb,Aaboud:2018pii}. The corresponding lower bounds on the Nelson-Barr fermions are 
\begin{align}
&m_T\simeq (\mu^2+\vert \vec{B}^u\vert^2)^{1/2}\gtrsim 1.4\text{ TeV}\, ,\\
&m_B\simeq(\mu^2+\vert \vec{B}^d\vert^2)^{1/2}\gtrsim 1.2\text{ TeV}\, .
\end{align}
Notice that these bounds have some degree of model dependence and can be in principle relaxed by modifying the dominant branching ratios of the heavy fermions. Given the mild bounds from flavor in this setup it would be interesting to pursue this way but we will stick to the simplest scenario and comply with the bounds above in what follows. 

\subsection{The Nelson-Barr quality problem}\label{sec:NBquality}

We can summarize the features of the BBP model discussed as follows:

First, The smallness of the strong CP phase is guaranteed by the spontaneous breaking of CP at the scale $f$ and by a non-anomalous $\mathbb{Z}_{2}^{\text{NB}}$-symmetry which forbids $\theta_{\text{QCD}}$ to be generated at tree-level. 

Second, Eq.~\eqref{eq:cond1} and Eq.~\eqref{eq:CPviol} guarantees the CKM phase to be order one. This conditions together with the condition in Eq.~\eqref{eq:quantumcorr} to ensure a small enough $\theta_{\text{QCD}}$, after quantum corrections are taken into account, requires the scale of CP breaking to be high enough. In particular assuming $\mu\simeq \vert\vec{B}^u\vert$ and applying the  bounds in Sec.~\eqref{sec:flavorconst} we get 
\begin{equation}
f_{\text{CP}}\gtrsim 1.4\cdot10^{3}\text{ TeV}\cdot\left(\frac{10^{-3}}{\vert \vec{k}^u\vert}\right)\cdot\left(\frac{\mu}{1.5\text{ TeV}}\right)\, . \label{eq:upperboundonf}
\end{equation}

To quantify the quality problem of our theory, let us assume that our model is valid up to a cut-off scale $\Lambda$. Thus, we expect higher dimensional operators to be generated, suppressed by this cutoff. These can affect the Nelson-Barr texture resulting in finite threshold corrections to $\theta_{\text{QCD}}$ like the ones discussed in Sec.~\ref{sec:radcorr}.  Since  $\mathbb{Z}_{2}^{\text{NB}}$ is non-anomalous we can assume the high energy theory to respect this symmetry (possibly lifted to a non-anomalous gauge symmetry in the UV spontaneously broken to a discrete subgroup in the IR). The first dangerous operators are then 
\begin{equation}
\mathcal{L}_{\mathbb{Z}_{2}^{\text{NB}}-\text{even}}\supset c^{q}_i \frac{\Sigma HQ_i\tilde{U}}{\Lambda}+c^\mu\frac{\Sigma^2U\tilde{U}}{\Lambda}+\dots\label{eq:higher}
\end{equation}
where we include only the $\mathbb{Z}_{2}^{\text{NB}}$-even operators arising at the leading order in $\Sigma/\Lambda$. 

The first operator is particularly dangerous, looking at Eq.~\eqref{eq:deltatheta2} we see that if we take $\vert c^{q}_i\vert\sim\mathcal{O}(1)$ for every $i$, requiring the correction to the strong CP phase to be smaller than its present upper bound implies  

\begin{equation}
c_1^q\frac{f_{\text{CP}}}{\Lambda}\lesssim 10^{-15} \cdot\left(\frac{y_\text{up}}{10^{-5}}\right) \ ,\qquad   c^\mu\frac{f_{\text{CP}}}{\Lambda}\lesssim 10^{-13}\cdot\left(\frac{\vert \vec{k}^u\vert}{10^{-3}}\right)\, ,                      \label{eq:qualitybounds}
\end{equation}  
where the correction controlled by $c_1^q$ is enhanced by the inverse of the up Yukawa unless a particular texture is assumed in the $c^q_i$ coefficients. The second correction gets also enhanced by $f_{\text{CP}}/\mu\simeq 1/\vert \vec{k}^u\vert\gtrsim 10^3$. Using the upper bound on the scale of CP breaking derived in Eq.~\eqref{eq:upperboundonf} we find:\footnote{We used the reduced Planck mass, $M_{\text{Pl}}=2.4\cdot10^{18}\text{ GeV}$.}
\begin{align}
	&\Lambda\gtrsim 10^{3} M_{\text{Pl}} \cdot\left(\frac{c^q_1}{1}\right) \cdot\left(\frac{10^{-3}}{\vert \vec{k}^u\vert}\right)\cdot\left(\frac{\mu}{1.5\text{ TeV}}\right) \label{eq:cut1}\, , \\
	&\Lambda\gtrsim 10\, M_{\text{Pl}} \cdot\left(\frac{c^\mu}{1}\right) \cdot\left(\frac{10^{-3}}{\vert \vec{k}^u\vert}\right)^2\cdot\left(\frac{\mu}{1.5\text{ TeV}}\right)\, ,                               \label{eq:cut2}
\end{align}  
which show how the cut-off needs to be at the Plank scale for $c^\mu\simeq\mathcal{O}(1)$ and even higher for $c^q_1\simeq\mathcal{O}(1)$.

This implies that even accepting the fine tuning of the mass scale of fundamental scalar $\Sigma$, the Nelson-Barr mechanism in its simplest implementation suffers from a \emph{quality problem}: the NB texture is spoiled by Planck suppressed operators unless their Wilson coefficients are accidentally small. Of course the problem is exacerbated if the cut-off scale $\Lambda$ is lower than $M_{\text{Pl}}$ as showed in Eq.~\eqref{eq:qualitybounds}.\footnote{The quality problem exhibited here can in principle be ameliorated by introducing an extra symmetry acting on one heavy chiral fermion $U$ and the singlet $\Sigma$. However, other states should be added to the theory in order to make this symmetry non-anomalous, leading to non-minimal Nelson-Barr constructions.} 

An easy way to ameliorate this issue is to assume the UV physics to communicate only to the $U \tilde{u}$ bi-linear. In such a theory the Wilson coefficients of the operators in Eq.~\eqref{eq:higher} will be further suppressed as $c^q_1\simeq c^\mu\sim \mu/\Lambda$ because a $\mu$-insertion is required in order to break the $Z_2^{\text{VL}}$ symmetry under which $\tilde{U}$ is odd and all the other fields in the theory are even. The upper bound on the cut-off scale becomes then less severe
\begin{align}
&\Lambda\gtrsim  10^9\text{ TeV} \cdot\left(\frac{c^q_1}{1}\right)^{1/2} \cdot\left(\frac{10^{-3}}{\kappa}\right)^{1/2}\cdot\left(\frac{\mu}{1.5\text{ TeV}}\right) \label{eq:cut1good}\ , \\
&\Lambda\gtrsim 10^8\text{ TeV}\cdot\left(\frac{c^\mu}{1}\right)^{1/2} \cdot\left(\frac{10^{-3}}{\kappa}\right)\cdot\left(\frac{\mu}{1.5\text{ TeV}}\right)\ .                             \label{eq:cut2good}
\end{align}  
As the $Z_2^{\text{VL}}$-symmetry is clearly anomalous (only single chiral field $\tilde{U}$ is charged under it) we do not expect this symmetry to be respected by a UV theory of gravity. Therefore, one still has to estimate the impact of operators like the one in Eq.~\eqref{eq:higher} but suppressed by the Planck scale. Substituting $\Lambda=M_{\text{Pl}}$ in Eq.~\eqref{eq:cut1} and Eq.~\eqref{eq:cut2} one can see that they will invalidate the BBP construction, unless $c^q_1<10^{-3}$ and $c^{\mu}<0.1$. 

In what follows we will exhibit a UV completion of the BBP model where the scale of the singlet VEV $\Sigma$ is stabilized by dimensional transmutation, the same UV completion also allows us to alleviate the NB quality problem discussed here. The scaling in Eq.~\eqref{eq:cut1good} and Eq.~\eqref{eq:cut2good} will be explicitly realized in our UV completion and the effect of Planck suppressed operators is significantly reduced.

\section{A Nelson-Barr model from a hidden QCD at $\theta'=\pi$}\label{sec:UVcompletions}

\begin{figure}[h]
	\centering
	\includegraphics[width=0.8\linewidth]{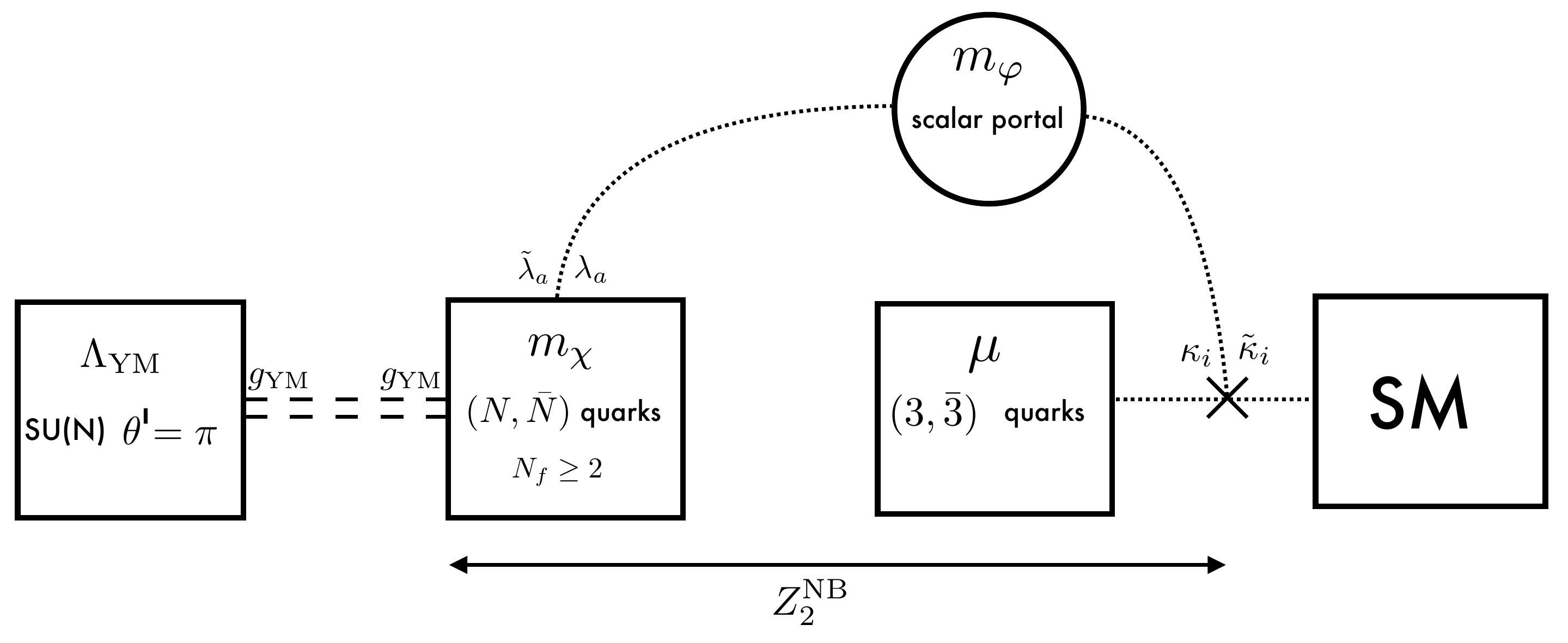}
	\caption{Cartoon of the Nelson Barr UV completion proposed here. The hidden sector is a $SU(N)$ Yang Mills theory at $\theta'=\pi$ with $N_f$ heavy flavors at the scale $m_\chi$. The scale of CP-breaking is related to the confinement scale the gauge theory: $f_{\text{CP}}\simeq \Lambda_{\text{QCD'}}$. Depending on the hierarchy between $m_\chi$ and $\Lambda_{\text{QCD'}}$ we have two scenarios: i) $m_\chi\lesssim\Lambda_{\text{QCD'}}$ with CP broken by the chiral condensate ii) $m_\chi\gtrsim\Lambda_{\text{QCD'}}$ with CP broken by the glueball condensate. We assumed that \emph{only} a scalar portal connects the hidden vector-like quarks with the light vector-like fermion pair at a much lower scale $v\lesssim\mu\ll m_\chi $. The latter mixes with the SM quarks like in the BBP model of Sec.~\ref{sec:BBP-Intro}.  The discrete symmetry ensuring the NB texture is a $Z_{2}^{\text{NB}}$ acting on the scalar portal and the new fermions. In order to treat the scalar portal perturbatively we take $m_\varphi\gg\Lambda_{\text{QCD'}}$.}
	\label{fig:the-overall-picture}
\end{figure}

We present here a UV completion of the NB mechanism where CP is broken dynamically by a hidden $SU(N)$ Yang Mills theory at $\theta'=\pi$, with $N_f$ hidden quarks. For simplicity, we take all hidden quarks to have same mass $m_\chi$. The hidden sector CP breaking is mediated to the visible sector by a scalar portal. The visible sector is assumed to have a structure similar to that of the BBP model of Sec.~\ref{sec:BBP-Intro}. As shown below, the hidden quarks are also charged under the $Z_{2}^{\text{NB}}$ and their mass, $m_\chi$, breaks this symmetry softly. Notice that for $N_f=2$ $Z_{2}^{\text{NB}}$ can be made non-anomalous. The schematics of our construction is illustrated in Fig.~\ref{fig:the-overall-picture}.

\subsection{Explicit models}\label{sec:models}
Here we present two explicit models which illustrate our basic idea. The first assumes the CP violation (CPV) is broken by a chiral condensate, while in the second model it is broken effectively by a pure YM theory at $\theta'=\pi$. One can intepulate between the two models by varying the hidden quarks' mass from $m_{\chi_1} \lsim\Lambda_{\text{QCD'}}$ to $m_{\chi_1} \gsim\Lambda_{\text{QCD'}}$. In Fig.~\ref{fig:parameter_space_of_the_model}, we show the viable parameter space of both these of models.
\paragraph{CP violation from hidden chiral condensate:} 
for $m_\chi\lesssim \Lambda_{\text{QCD'}}$ the CP breaking is driven by a chiral condensate,
 \begin{equation}
	\langle\chi_\ell\tilde{\chi}_\ell\rangle \sim\Lambda_{\text{QCD'}}^3\,e^{i\eta}\, , \label{eq:chiral_condanste}
	\end{equation} 
where $\chi_\ell\, ,\tilde{\chi}_\ell$ are vector-like pairs of hidden quarks, and $\ell$ stands for a flavor index. It is important to note that Eq.~\eqref{eq:chiral_condanste} is holds for $m_\chi\simeq \Lambda_{\text{QCD'}}$.\footnote{For $m_\chi\ll \Lambda_{\text{QCD'}}$, the imaginary part of $\langle\chi_i\tilde{\chi}_i\rangle$ is suppressed compare to its real part. As a result an order one CKM phase cannot be obtained.}
Consider a hidden $SU(N)$ Yang-Mills theory at $\theta'=\pi$ with $N_f\leq2$. CP can be spontaneously broken by the VEV of $\eta$ provided that~\cite{Dashen:1970et,DiVecchia:2017xpu,Witten:1980sp}  
\begin{equation}
\frac{1}{m^2_{\chi_1}}-\frac{f_{\eta}^2}{\chi_{YM}}<0\quad \Rightarrow\quad  \frac{\Lambda_{\text{QCD'}}^2}{N_\text{c}}\lsim m_{\chi_1}^2   \lsim\Lambda_{\text{QCD'}}^2 \, ,\label{eq:ineqchiral}
\end{equation}
with $N_\text{c}=N$ the number of colors, $m^2_{\chi_1}$ the mass of the heaviest generation, $f_{\eta}\sim \sqrt{N_c}\,\Lambda_{\text{QCD'}}$, while $\chi_{YM}\sim\Lambda_{\text{QCD'}}^4$ is independent on $N_c$, at leading order in the $1/N_\text{c}$-expansion. The inequality in Eq.~\eqref{eq:ineqchiral} sets a lower bound for the bare mass of the hidden quarks and can be easily generalized to the case of $N_f$ heavy flavors~\cite{Dashen:1970et,DiVecchia:2017xpu,Witten:1980sp}. 

The CP-phase in the chiral condensate is then mediated to the visible sector via four-fermi operators involving new vector-like fermions \`a la BBP. One way to generate these operators is to consider a scalar portal 
	\begin{align}
	 	\mathcal{L}_\text{scalar-portal} = & -\frac{1}{2}m_\varphi^2\varphi^2 -m_\chi \chi_\ell\tilde{\chi}_\ell-\mu U\tilde{U}
	 	\nonumber \\ 
	 	& +\left(\lambda\varphi+\tilde{\lambda}\varphi^{\dagger}\right)\chi_\ell\tilde{\chi}_\ell
	 	+\left(\kappa_{i}\varphi+\tilde{\kappa}_{i}\varphi^{\dagger}\right)\tilde{u}_{i}U+\text{h.c.} \label{eq:scalar.portal1} 
	\end{align}
where $\varphi $ is a heavy scalar messenger, $U,\tilde{U}$ are heavy vector-like up quarks, while $\tilde{u}_i$ are the three generations of the $SU(2)_L$ singlets of the would be SM anti-up fields.
To achieve the structure of Eq.~\eqref{eq:scalar.portal1}, we extend the $Z_2^{\text{NB}}$ to allow also the anti-hidden quarks to transform under it, 
\begin{equation}
\mathbb{Z}_2^{\text{NB}}: U\rightarrow -U\: , \: \tilde{U}\rightarrow -\tilde{U} \: , \: \Sigma\to -\Sigma \: , \: \chi_\ell \rightarrow +\chi_\ell \: , \: \tilde{\chi}_\ell \rightarrow - \tilde{\chi}_\ell \, .\label{eq:Z2NBchiral}
\end{equation}
By this construction, the mass $m_\chi$ is a source of soft $\mathbb{Z}_2^{\text{NB}}$ breaking in the model. For simplicity, we take the coupling $\lambda, \tilde{\lambda}$ to be diagonal in the quarks mass basis. Integrating out the heavy complex scalar, assuming $m_\varphi\gg \Lambda_{\text{QCD'}}$, we get 
\begin{equation}
	\mathcal{L}_\text{portal}\supset\left(\frac{\tilde{\lambda}\kappa_i \chi_\ell \tilde{\chi}_\ell+\lambda\tilde{\kappa}_i \chi_\ell^{\dagger}\tilde{\chi}_\ell^{\dagger}}{m_\varphi^2}\right)\tilde{u}_i U-\mu U \tilde{U}+\dots \label{eq:fourfermiok}
\end{equation}
where the dots includes higher dimensional operators, such as four fermi operators involving either the hidden quarks only or the visible quarks only.  
After the chiral symmetry breaking, the flavor vector $B^u_i$ in Eq.~\eqref{eq:defBu} scales as
\begin{equation}
\vert B^u_i\vert \sim\frac{\vert \lambda\kappa_i\vert\Lambda_{\text{QCD'}}^3}{ m_\varphi^2}\, ,\label{Buchiral}
\end{equation}
assuming $\vert k_i\vert \sim \vert\tilde{k}_i\vert$ and $\lambda\sim \tilde{\lambda}$. 
 An order one CKM phase can easily be obtained by requiring $B^u/\mu\gtrsim 1$, while the smalless of the NB Yukawa coupling is ensured even for $\kappa\simeq\mathcal{O}(1)$ by having $\Lambda_{\text{QCD'}}\ll m_\varphi$. 
 
Here we discuss the general challenges in our construction that lead to the parameter space in the of Fig.~\ref{fig:parameter_space_of_the_model}: 
\begin{itemize}
\item The $Z_2^{\text{NB}}$-even operators in Eq.~\eqref{eq:higher} are in principle generated from 1-loop radiative corrections with one insertion of the hidden fermion condensate and one $\mu$ insertion. However, the resulting wrong Yukawa is controlled by $(Y_u B^{u\ast})_i$ which does not lead to a new contribution to $\Delta\theta_{\text{QCD}}$ as can be seen from Eq.~\eqref{eq:deltatheta2}. Contribution to $\Delta\theta_{\text{QCD}}$ comes at higher loop order or additional insertions of $\langle\chi_\ell\tilde{\chi}_\ell\rangle $ and thus is highly suppressed.
\\
\item One of the problems of this NB UV completion, as pointed out in \cite{Dine:2015jga}, is that the UV fermion mass $m_{\chi} \tilde{\chi}_1\chi$ breaks the $\mathbb{Z}_2^{\text{NB}}$ softly. Since the masses cannot be too far from $\Lambda_{\text{QCD'}}$ in order for CP to be broken, suppressing the  $\mathbb{Z}_2^{\text{NB}}$-breaking operators requires $\Lambda_{\text{QCD'}}$ to be small compared to scalar portal mass $m_\varphi$.  For example, a real correction to the "wrong Yukawa" is proportional to $\mathbb{Z}_2^\text{NB}$ soft breaking scale $m_\chi$.

\item A related problem is that the soft-breaking of the $\mathbb{Z}_2^{\text{NB}}$ allows for tadpole contributions to the NB Yukawa, which are purely real. These contributions are unavoidable, and requiring them to be smaller than $B^u$ in Eq.~\eqref{Buchiral} gives a lower bound on the confinement scale
\begin{equation}
\Lambda_{\text{QCD'}}\gtrsim \frac{m_\chi}{(16\pi^2)^{1/3}}\, .
\end{equation}
For the chiral model, the above bound is always realized. However, as we discussed in the following, it is forms a challenge for the effective hidden YM variant of the model. We show how this challenge can be addressed by further model building.

\item Lastly, this chiral model can solves the strong CP problem, and at the same time lead to an order one CKM phase, only in a very narrow region of its parameter space. Therefore, in what follows we consider a different model, where the allowed parameter space is much larger. 
\end{itemize}

\subsubsection{CP violation from hidden glueball condensate:} For $m_\chi\gg\Lambda_{\text{QCD'}}$, the CP breaking is driven by the glueball condensate and we have 
\begin{equation}
\langle F^2+i F\tilde{F}\rangle\sim\Lambda_{\text{QCD'}}^4e^{i\eta}\ .
\end{equation}
The hidden sector breaks CP spontaneously by effectively a pure glue Yang-Mills theory. Recent results in field theory suggest that this theory breaks CP spontaneously at $\theta'=\pi$ even at finite $N_c$ \cite{Gaiotto:2017yup,Gaiotto:2017tne}, generating an unsuppressed CP phase which is the necessary starting point of any NB construction.\footnote{ At large $N_c$ the CP violating phase will always be suppressed by $1/N_c$ as reviewed in Appendix~\ref{app:spontaneousCP}. In the same Appendix we review the arguments of~\cite{Gaiotto:2017yup,Gaiotto:2017tne} suggesting that the well known large $N_c$ result remains valid at finite $N_c$.}
The CP-phase can be mediated to the visible sector via a scalar portal similar to the one in eq.~\eqref{eq:scalar.portal1} after integrating out the heavy hidden fermions. Assuming $m_\phi\gg m_\chi$, integrating out the heavy scalar leads to four fermi operators of the same form of Eq.~\eqref{eq:fourfermiok}. Then, for $m_\chi\gg \Lambda_{\text{QCD'}}$, we can integrate out the fermions perturbatively.  The resulting dimension seven portal can be written as 
\begin{equation}\label{eq:Gluevall_portal}
\mathcal{L}_\text{portal}\supset\frac{1}{m_\varphi^2 m_\chi}\frac{\alpha'_sN_f}{8\pi} \left(\frac{g_i+\tilde{g}_i}{2}F^2+\frac{g_i-\tilde{g}_i}{2} i F\tilde{F}\right)\tilde{u}_i U - \mu U \tilde{U}+\dots\ .
\end{equation}
where we defined $g_i = \lambda \tilde{\kappa}_i+\tilde{\lambda} \kappa_i$ and $\tilde{g_i} = \tilde{\lambda} \tilde{\kappa}_i+\lambda \kappa_i$ in the same notation of Eq.~\eqref{eq:scalar.portal1}. After confinement, the flavor vector $B^u_i$ in Eq.~\eqref{eq:defBu} scales as 
\begin{equation}
\vert B^u\vert \sim\frac{\alpha'_s\vert g_i\vert N_f}{8\pi} \frac{\Lambda_{\text{QCD'}}^4}{m_\varphi^2 m_\chi}\, , \label{eq:B_pure_gloun}  
\end{equation}
where we assumed $g_i\sim \tilde{g}_i$. 
 Similarly to the chiral condensate model, an order one CKM requires $B^u/\mu\gtrsim 1$, while the suppression of the NB yukawa coupling is ensured by having $\Lambda_{\text{QCD'}}\ll m_\varphi$. 

\begin{figure}[h!]
	\begin{centering}
		\includegraphics[scale=0.36]{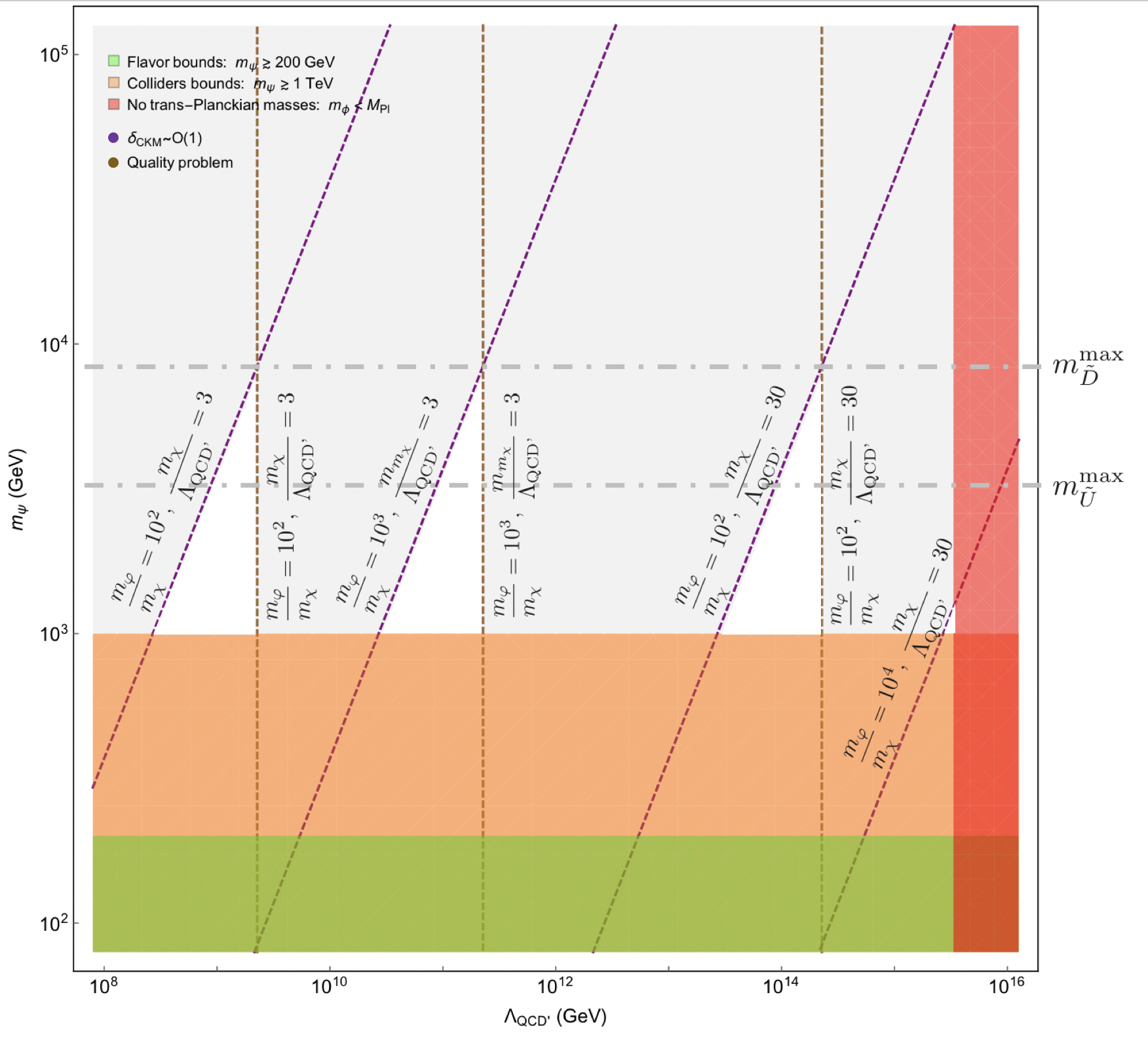}\tabularnewline
			\caption{The allowed parameters space of the model for different hierarchy between $m_\varphi,m_\chi$ and $\Lambda_{\text{QCD'}}$. The colored regions are excluded regions as written in the legend. The \textcolor{LimeGreen}{green} and the \textcolor{orange}{orange} regions are bounded by flavor and colliders bounds on the mass of a heavy vector like down quarks, as explained in section~\ref{sec:flavorconst}. The \textcolor{red}{red} region is excluded by transplanckian mass scales. The region to the left of each \textcolor{RedViolet}{purple} contour line, i.e. for each choice of $m_\varphi,m_\chi$, is excluded by Eq.~\eqref{eq:cond1}, which guarantees a large CKM phase. Similarly, the region to the right of the \textcolor{brown}{brown} vertical contour lines is excluded if we allow dimension 5 Planck suppressed operator with an order one coupling as in Eq.~\eqref{eq:higher}. The two dashed \textcolor{gray}{gray} horizontal lines are the maximal mass allowed to new add heavy vector-like quarks if the down sector: $m_{\tilde{D}}$ and the up sector: $m_{\tilde{U}}$.	
			In a more relaxed quality problem scenario, the \textcolor{brown}{brown} vertical contour lines can be ignored. This can be achieved by gauging the discrete symmetries, thus allowing only higher dimensional operators in the theory, or by setting the coupling of the operator in Eq.~\eqref{eq:higher} to be much smaller than one. In this relaxed scenario, the upper bound on $m_{\tilde{D}}$ and $m_{\tilde{U}}$ is also relaxed and can take much higher values.	
		}
		\label{fig:parameter_space_of_the_model}
		\par \end{centering}
\end{figure}

Interestingly, the challenges of this model are quite different than the related to the chiral condensate, essentially for two reasons: i) the VEV of the CP-breaking operator does not break $Z_2^{\text{NB}}$ spontaneously here. The only source of $Z_2^{\text{NB}}$ breaking is the soft breaking controlled by the hidden fermions masses.  Since the hidden fermion scale is heavy, $Z_2^{\text{NB}}$ is badly broken in the IR and there is no real distinction between the $Z_2$-even and the $Z_2$-odd contribution ii) Since the scale of soft $Z^{\text{NB}}_2$-breaking is pushed to be higher than $\Lambda_{\text{QCD'}}$, the radiative stability of the NB construction is in danger. We will see how a viable model can still be achieved with extra structure, leading to the parameter space in the of Fig.~\ref{fig:parameter_space_of_the_model}.

\begin{figure}[h!]
	\begin{centering}
		\includegraphics[scale=0.2]{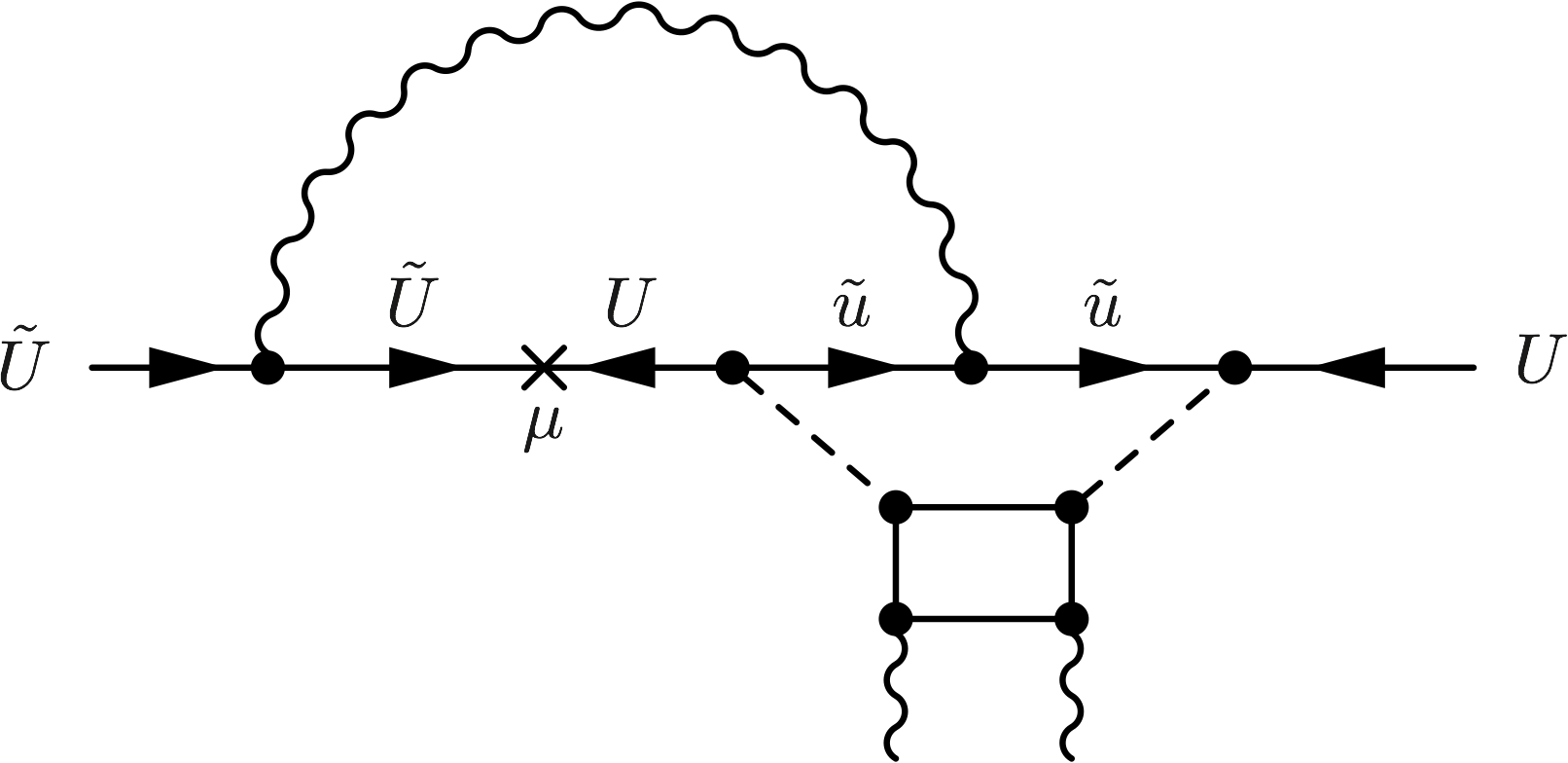}
		\caption{Imaginary contribution to the mass of the vector-like quarks due to the $\mathbb{Z}_2^{\text{NB}}$ soft breaking mass, $m_\chi$.}
		\label{fig:Diagrams_for_Our_NB_Model}
	\end{centering}
\end{figure}

The corrections to $\bar{\theta}_{\text{QCD}}$ can be calculated by spurion analysis, as seen by Eqs.~\eqref{eq:deltatheta1}-\eqref{eq:deltatheta3}. For example, in the BBP model~\cite{Bento:1991ez}, the correction to $\bar{\theta}_{\text{QCD}}$ are proportional to $\Sigma^2$, where $\Sigma$ is the scalar which breaks both $\mathbb{Z}_2^\text{NB}$ and CP spontaneously. In our model, CP is broken by the VEV of effective dimension 4 operator such as $F\wedge F$. Thus, one may relate the CP violation scale in the BBP model, to our model by

\begin{equation}
\left\langle \varphi\right\rangle \sim \left\langle \Sigma\right\rangle \longleftrightarrow\frac{g_{s'}^2\;\Lambda_{\text{QCD'}}^{4}}{16\pi^{2}m_{\chi}m_{\varphi}^{2}}\,.
\end{equation}
where $g_{s'}$ is the hidden SU(N) YM couplings, which is evaluated at the strong coupling scale.  
However, as in this model the CPV operator does not break the $\mathbb{Z}_2^\text{NB}$ symmetry, we expect additional contributions to $\bar{\theta}_{\text{QCD}}$ that are linear in $\Sigma$ as illustrated in Fig~\ref{fig:Diagrams_for_Our_NB_Model}. 

\begin{figure}[h!]
	\begin{centering}
		\includegraphics[scale=0.17]{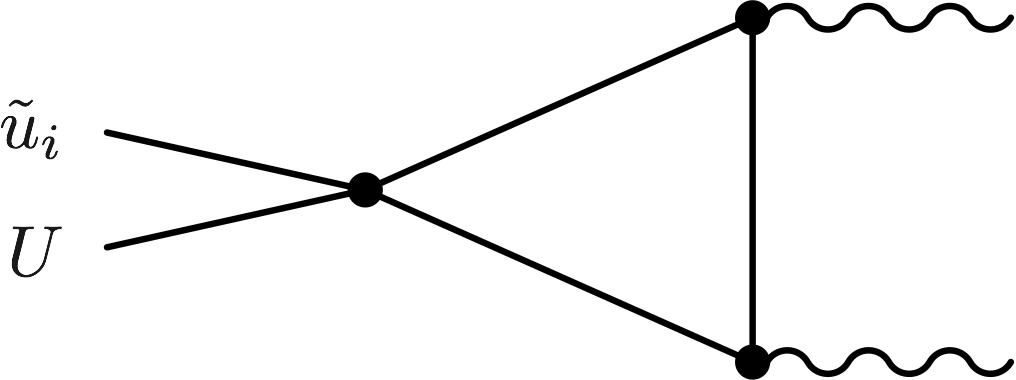}\qquad \qquad\qquad		\includegraphics[scale=0.2]{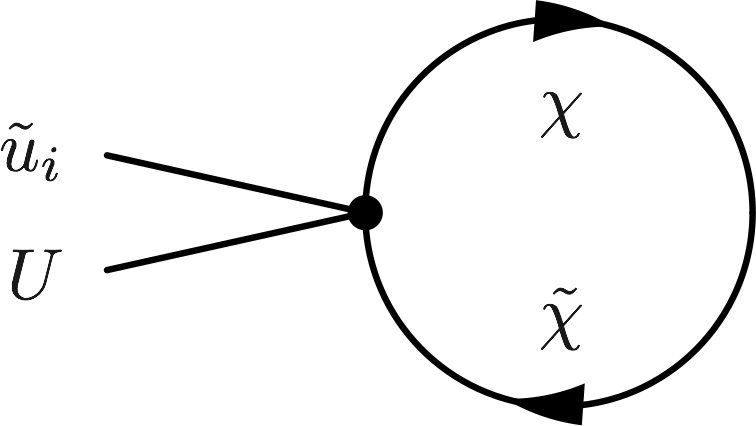}	
		\caption{{\bf left:} complex contribution of the gluon condensate to the NB coupling in  Eq.\eqref{eq:EFT_NB_part_below_confinement}.\qquad  { \bf right:} only real contribution to the NB coupling from the tadpole.}	
		\label{fig:tadpole_glueball_model}	
		\par \end{centering}
\end{figure}

Considering radiative corrections from the UV Lagrangian, we can estimate the threshold contributions to $\bar{\theta}_{\textrm{QCD}}$ by spurion analysis up to a loop suppression which depends on the different corrections. This depends on the dimensionfull scales as well as the symmetries of the model. A similar spurion analysis can be found in \cite{Davidi:2017aa}. We find that the leading contribution comes at order
\begin{align}
\delta\bar{\theta}_{\text{rad}}\simeq \frac{g_{s'}^2\;\Lambda_{\text{QCD'}}^{4}}{(16\pi^{2})^3m_{\chi}^{2}m_{\varphi}^{2}}<10^{-10}\,.\label{eq:Radiative_correction_Bound}
\end{align}
The above constraint of Eq.~\eqref{eq:Radiative_correction_Bound}
is easily satisfied as the set up of this model includes $\Lambda_{\text{QCD'}}<m_{\chi}<m_{\varphi}$.
As a result, in a composite UV completion, the small portal coupling is automatically achieved from the hierarchy between the scale of the portal and the dynamical scale of CP breaking. 
\paragraph{The tadpoles challenge:} As already discussed in the context of the chiral NB model, their are dangarous contributions coming from tadpoles diagrams. Integrating out the hidden quarks from the operator $\chi{}_\ell\tilde{\chi}_\ell\tilde{u}_{i}U$, resulting in an effective coupling between $\tilde{u}_{i}U$ and the SU(N) gauge bosons as in \eqref{eq:L_NB2}. This is  illustrated by the diagrams of Fig.~\ref{fig:tadpole_glueball_model}.
\\
The right diagram of Fig.~\ref{fig:tadpole_glueball_model} gives a threshold correction to the mass of $U\tilde{u}$ of the order of
\begin{equation}
\L^{\text{EFT}}_{\text{Tadpole}} \supset  \frac{g_{i}+\tilde{g}_{i}}{16\pi^2 m_{\varphi}^{2}} N_{f}m_\chi^{3}\log\left(\frac{m_\chi^{2}}{\Lambda^{2}}\right)\tilde{u}_{i}U +\text{h.c.}\, . \label{eq:EFT_NB_part_Tadpole}
\end{equation}
In comparison, the left diagram of Fig.~\ref{fig:tadpole_glueball_model} gives a threshold correction to the effective coupling of the order of,
\begin{equation} \label{eq:L_NB2}
\L^{\text{EFT}}_{\text{Gluons}} \supset  \frac{\alpha_{s}'\,N_{f}}{2\pi m_{\varphi}^{2}m_{\chi}}\left(\frac{g_{i}+\tilde{g}_{i}}{2}F_{\,\mu\nu}^{n}F^{n\,\mu\nu}+i\frac{g_{i}-\tilde{g}_{i}}{2}F_{\,\mu\nu}^{n}\tilde{F}^{n\,\mu\nu}\right)\tilde{u}_{i}U+\text{h.c.} \, .
\end{equation}
Below the SU(N) confinement scale, the effective couplings between $\tilde{u}_iU$ and the hidden gluons gives the desirable CPV mass mixing
\begin{equation} 
\L^{\text{EFT}}_{\text{Gluons}} \longrightarrow \frac{g_{s'}^2\; \Lambda_{\text{QCD'}}^{4} N_{f}}{16\pi^{2}m_{\chi}m_{\varphi}^{2}} \left(g_{i}e^{i\eta}+\tilde{g}_{i}e^{-i\eta}\right)\tilde{u}_{i}U+\text{h.c.} \, , \label{eq:EFT_NB_part_below_confinement}
\end{equation} 
where $\eta\sim\mathcal{O}(1)$ if there is no fine tuned alignment between $g_i$ and $\tilde{g}_i$. \\
In order to achieve an order one CKM phase of the low energy SM effective theory, and a viable NB model, the total phase in the effective $B_i\tilde{u}_iU$ term should not be suppressed. Therefore, the correction from \eqref{eq:EFT_NB_part_Tadpole} should be smaller or equal to the one from \eqref{eq:EFT_NB_part_below_confinement}. This result in a catastrophic bound,
\begin{equation}
g_{s'}^2\Lambda_{\text{QCD'}}^{4} \gtrsim m_\chi^4 \,. \label{eq:Tadpole_bound}
\end{equation} 
Since we assume $\Lambda_{\text{QCD'}}<m_\chi$, the bound of \eqref{eq:Tadpole_bound} is not easily satisfied. Without further model building, the theory if vailabe  if we assume a very strong coupling $g_s > 1$, but even in this limit, the hidden quarks mass should be just above the confinement scale:
\begin{equation}
m_{\chi}\lesssim \sqrt{4\pi}\,\Lambda_{\text{QCD'}}\;.
\end{equation}
In order to have a more robust separation between $m_\chi$ and $\Lambda_{\text{QCD'}}$, we need to invoke further construction to our model. For example, we can ameliorate the real tadpole contribution compare to the CPV contribution from the hidden gluons, by using a mechanism similar to the Twin Higgs model~\cite{Chacko:2006aa,De_Simone_2013}. In contrast to the original Twin Higgs model, we are not trying to solve the Higgs hierarchy problem. Thus, we do not add additional scalars nor we treat the Higgs or any other scalar as a pseudo Goldstone boson.
We introduce a second hidden sector of quarks, which are charged by another $SU(N_B)$ gauge group. For convenience, we defined our original hidden sector by sector-$A$ and the additional hidden sector by sector-$B$. Since the scalar mediator is not charged under any of the $SU(N_A)$ or $SU(N_B)$ gauge group, it considered to have a similar Yukawa coupling to the hidden sector-B quarks.
\begin{align*}
\L^A_{\text{Hidden}}= & -\frac{1}{4g_{A}^{2}}\text{Tr}F^2_{A}+\frac{\theta_{A}}{8\pi^{2}}\text{Tr}F_{A}\tilde{ F}_{A}+i\bar{\chi}_{A}\cancel{D}\chi_{A}+i\bar{\tilde{\chi}}_{A}\cancel{D}\tilde{\chi}_{A}\\
& -m_{\chi}\chi_{A}\tilde{\chi}_{A}+\left[\lambda\varphi+\tilde{\lambda}\varphi^{*}\right]\chi_{A}\tilde{\chi}_{A}\\
\\
\L^B_{\text{Hidden}}= & -\frac{1}{4g_{B}^{2}}\text{Tr}F^2_{B}+\frac{\theta_{B}}{8\pi^{2}}\text{Tr}F_{B}\tilde{ F}_{B}+i\bar{\chi}_{B}\cancel{D}\chi_{B}+i\bar{\tilde{\chi}}_{B}\cancel{D}\tilde{\chi}_{B}\\
& -m_{\chi}\chi_{B}\tilde{\chi}_{B}+\left[-\lambda\varphi-\tilde{\lambda}\varphi^{*}\right]\chi_{B}\tilde{\chi}_{B}
\end{align*}
The above form of the Lagrangian is achieved by imposing the exchange Twin symmetry:
\begin{align}
{\cal P}:\quad\chi_{A} & \longleftrightarrow  \chi_{B} \nonumber\\
\tilde{\chi}_{A} & \longleftrightarrow  \tilde{\chi}_{B} \nonumber\\
F_{A} & \longleftrightarrow  F_{B}\nonumber \\
\varphi & \longleftrightarrow  -\varphi \label{eq:Twin_exchange_transformation}
\end{align}
To first loop order, the tadpole correction from $\chi_{A}\tilde{\chi}_{A}$ as shown in the right diagram of Fig~\ref{fig:tadpole_glueball_model}, is linear in $+\lambda$ and independent of $g_A$. In contrast, in the same order the tadpole correction from $\chi_{B}\tilde{\chi}_{B}$ is linear in $-\lambda$  and the resulting tadpole contribution of \eqref{eq:EFT_NB_part_Tadpole} vanishes.
The next loop order scales as $\frac{m_{\chi}^{3}}{16\pi^{2}}\left(\frac{g_{A}^{2}-g_{B}^{2}}{16\pi^{2}}\right)\,$. In the limit $g_{A}\rightarrow g_{B}$ this second loop order also vanish. In fact, the Twin exchange symmetry is considered to be an "exact" symmetry only in the limit of $g_{A}=g_{B}$. Indeed, the value of $g^2_{A}-g^2_{B}$ can be very small, but we have no reason to believe it vanishes identically. Suppressing the tadpole contribution looks promising but it is not enough to solve the problem. The additional sector also modify the low energy CKM phase. To have a viable solution we must require that the complex parameter in the mass mixing, $B_i\tilde{u}_iU$, should not be suppressed. By power counting we find
\begin{equation}
B_{i}  \simeq\frac{|\lambda| m_{\chi}^{3}}{16\pi^{2}}\left(\frac{g_{A}^{2}-g_{B}^{2}}{16\pi^{2}}\right)+\frac{|\lambda|}{m_{\chi}}\left[\frac{g_{A}^{2}\left(\left\langle F_{A}F_{A}\right\rangle +i\left\langle F_{A}\tilde{F}_{A}\right\rangle_{\theta_A} \right)-g_{B}^{2}\left(\left\langle F_{B}F_{B}\right\rangle +i\left\langle F_{B}\tilde{F}_{B}\right\rangle_{\theta_B} \right)}{16\pi^{2}}\right] \label{eq:B_Twin_solution}
\end{equation}
where for simplicity we take $\lambda\sim\tilde{\lambda}\sim|\lambda|$.\\
Therefore, if we consider $\theta_A=\theta_B=\pi$, we notice that for $|g_{A}^{2}-g_{B}^{2}|\ll1$, we are back to the small corner of parameter space since the imaginary part of \eqref{eq:B_Twin_solution} is also suppressed. However, if we consider a "twisted" twin sector, where $\theta_A =\pi$ while $\theta_B=0$, we find that the imaginary part of \eqref{eq:B_Twin_solution} is independent of $g_{B}$  as the VEV $i\left\langle F_{B}\tilde{F}_{B}\right\rangle_{\theta_B=0}=0$. In this scenario, one can consider $m_\chi\gg \Lambda_{\text{QCD'}}$ if one also require $|g_{A}^{2}-g_{B}^{2}|\ll1$ such that the CKM phase is not suppressed.

\paragraph{In summary} in Fig.~\ref{fig:parameter_space_of_the_model} we show the allowed parameter space of the model, which solves the strong CP problem and does not suffer from a quality problem. Flavor and collider bounds are shown in horizontal green and orange exclusions areas. The requirement of having order one CKM, Eq.~\eqref{eq:cond1}, is translated in our model into
\begin{equation}\label{eq:cond1_our}
m_\psi\lsim\vert B^{(u/d)}\vert \sim\frac{\alpha'_s\vert g_i\vert N_f}{8\pi} \frac{\Lambda_{\text{QCD'}}^4}{m_\varphi^2 m_\chi}\, ,
\end{equation}
where $m_\psi$ is the mass of the new heavy vector-like SM  up or down quarks. The new heavy vector-like mass can be taken to be up-like or down-like quarks, depending on the choice of the model. Therefore, for a fixed hierarchy between $\Lambda_{\text{QCD'}},m_\chi$ and $m_\varphi$, Eq.~\eqref{eq:cond1_our} creates a linear line in Fig.~\ref{fig:parameter_space_of_the_model}, which set the allowed parameter space for $m_\psi$. 

In addition, consider the impact of the following dimension five, Planck suppressed operator, $\frac{c_{_{\text{Pl}}}}{M_{\text{Pl}}}\varphi QH\tilde{u}$ for an up-sector NB model, or the operator $\frac{c_{_{\text{Pl}}}}{M_{\text{Pl}}}\varphi QH\tilde{d}$ for an down-sector NB model. These operators give rise to the NB quality problem. A solution to the quality problem, as presented in section~\ref{sec:NBquality}, is translated in our model into the bound
\begin{equation}\label{eq:cond2_our}
\vert B^{(u/d)}\vert < 10^{-10}  M_{\text{Pl}}  \frac{y_u \text{ or } y_d}{c_{_{\text{Pl}}}}\,.
\end{equation}
Note, in the above equation, $\vert B^{(u/d)}\vert$ is independent of $m_\psi$, as explained in Eq.~\eqref{eq:B_pure_gloun}. Thus, Eq.~\eqref{eq:cond2_our} creates the horizontal lines in Fig.~\ref{fig:parameter_space_of_the_model}, for the allowed parameter space as function of $\Lambda_{\text{QCD'}},m_\chi$ and $m_\varphi$. 

Moreover, combining Eq.~\eqref{eq:cond1_our} and Eq.~\eqref{eq:cond2_our}, we find different triangles, with almost exact area, of allowed parameter space for different choice of hierarchy between $\Lambda_{\text{QCD'}},m_\chi$ and $m_\varphi$. Finally, Eq.~\eqref{eq:cond1_our} and Eq.~\eqref{eq:cond2_our} gives an upper bound on the mass of the new SM vector-like quarks:
\begin{equation}\label{eq:upper_bound_vectorlike_mass}
m_{\tilde{U}}\lesssim 3.6\times 10^{3}\, \text{GeV}\, ,  \quad  \qquad m_{\tilde{D}}\lesssim 8.4\times 10^{3}\, \text{GeV}\, .
\end{equation}
\newpage

\section{Conclusions}\label{sec:conclusions}
We present a simple and calculable scheme, where spontaneous CP violation is induced by a hidden $\theta'=\pi$ QCD like dynamics, as suggested by recent theoretical developments~\cite{Gaiotto:2017tne,Gaiotto:2017yup}. The hidden sector is connected to the standard model (SM) Lagrangian through a scalar portal, such that the low energy effective theory reproduces the SM Lagrangian with a negligibly small strong CP phase. In this construction, the quality problem is avoided as the operator that breaks CP is composite, of mass dimension four. Our successful models requires a sizable hierarchy between the scale of the visible sector vector like quarks ($m_{\psi} \leq 10^4 \, \text{GeV}$), and the scale of hidden QCD like dynamics ($\Lambda_{\text{QCD'}} \geq 10^8\, \text{GeV}$). Thus, our model can be tested via future colliders and possibly by next generation neutron-electric-dipole-moment experiments. Finally, our model predicts a cosmological phase transition at temperatures of the order of the hidden sector confinement scale, in addition to a CP phase transition, possibly at even higher temperatures~\cite{Gaiotto:2017yup}.

\section*{Acknowledgments}
We are grateful to Diego Redigolo for many discussions and being our collaborator on this project till its last stage of preparations. We thank Zohar Komargodski for collaboration at the early stages of this paper and feedback. We also thank Michael Dine, Ryosuke Sato for fruitful discussions. The work of GP is supported by grants from The U.S.-Israel Binational Science Foundation (BSF), Israel Science Foundation (ISF), Friedrich Wilhelm Bessel research award, Minerva, Yeda-Sela-SABRA-WRC, and the Segre Research Award.
\appendix

\section{CP breaking at $\theta=\pi$ in pure Yang Mills}\label{app:spontaneousCP}
In this appendix we summarize the theoretical arguments showing that pure $SU(N)$ Yang Mills at $\theta=\pi$ breaks CP spontaneously. We start by presenting the large $N$ arguments~\cite{HOOFT1974461}, supported by explicit AdS/CFT construction in Type II	A string theory~\cite{Maldacena:1997re}. We then move to the more recent argument at finite $N$~\cite{Gaiotto:2017yup,Gaiotto:2017tne}, which is crucial to ensure an unsuppressed CP phase in our construction. 

\paragraph{Large N computation}
Consider the pure YM Lagrangian including a $\theta$-term,
\begin{align}
\mathcal{L}_{\text{YM}} & =-\frac{1}{2g^{2}_{\text{YM}}}\text{tr}(F^{\mu\nu}F_{\mu\nu})+\frac{\theta}{16\pi^{2}}\text{tr}(F^{\mu\nu} \tilde{F}_{\mu\nu})\\
& =N\left(-\frac{1}{2\lambda_N}\text{tr}F^{\mu\nu}F_{\mu\nu}+\frac{\theta}{16\pi^{2}N}\text{tr}F^{\mu\nu}\tilde{F}_{\mu\nu}\right) \label{eq:Large_N_Lagrangian}
\end{align}
where $ \tilde{F}_{\mu\nu}=  \epsilon_{\mu\nu\rho\sigma}F^{\rho\sigma}$ and we defined the \textquoteright t Hooft coupling,
\begin{equation}
\lambda_N\equiv g^{2}N\,.
\end{equation}
We consider the theory in the limit $N\rightarrow\infty$, with both $\Lambda_{\text{UV}}$ and $\lambda_N$ held fixed. By dimensional transmutation, the dynamical scale is given at 1-loop order by
\begin{equation}
\Lambda_{\text{QCD}}=\Lambda_{\text{UV}}\exp\left(-\frac{3}{22}\frac{\left(4\pi\right)^{2}}{g_{\text{UV}}^{2}N}\right)\,.
\end{equation} 
In this limit, the physical scale $\Lambda_{\text{QCD}}$ is also remains fixed.

In the basis where the factor $N$ multiply the entire Lagrangian of Eq.~\eqref{eq:Large_N_Lagrangian}, the $\theta$-term appears to be suppressed by 1/N compare to the YM kinetic term. At leading order in the 1/N expansion, the $\theta$ contributions to any physical observer is suppressed by $\frac{\theta}{N}$. The ground state energy, can be defined schematically in the Euclidean path integral by,
\begin{equation}
e^{-VE\left(\theta\right)}=\int{\cal D}A\exp\left(-\int d^{4}x\L_{\text{YM}}\right)\,,
\end{equation}
where $V$ is the spacetime volume. 

Large N arguments suggest that $E\sim N^{2}$. We assume that the $\theta$ term affects the energy levels as follows,
\begin{equation}
E\left(\theta\right)=N^{2}f\left(\frac{\theta}{N}\right)
\end{equation}\
for some function $f(x)$.

Moreover, we assume that the $\theta$ dependence remain in the large N limit for several reasons. First, if we consider the theory to include light quarks, as in QCD, by using chiral rotation, one can rotate the $\theta$ coupling from the topological term to the quark mass matrix. Thus, the dependence can be seen in the chiral Lagrangian.
Second, at leading order in the 1/N expansion, one should sum an infinite number of diagrams. Even if each diagram separately vanish in the large N limit, their infinite series might not.

For example, we can introduce the so called topological susceptibility,
\begin{equation}
\chi\left(k\right)=\int d^{4}xe^{ik\cdot x}\left\langle \text{tr}\left(F_{\mu\nu} \tilde{F}^{\mu\nu}\left(x\right)\right)\text{tr}\left(F_{\mu\nu} \tilde{F}^{\mu\nu}\left(0\right)\right)\right\rangle \,.
\end{equation}
This quantity responds to variations in $\theta$. Moreover, the ground state energy $E(\theta)$
has the dependence
\begin{equation}
\frac{d^{2}E}{d\theta^{2}} \propto \left(\frac{1}{16\pi^{2}N}\right)^{2}\lim_{k\rightarrow0}\chi\left(k\right)\,.
\end{equation}
One finds that, at leading order in 1/N, each individual diagrammatic correction to the topological susceptibility
satisfy $\chi\left(k\right)\rightarrow0$ as $k\rightarrow0$. However, the sum of all such diagrams may not vanish.
The energy which must satisfy $E(\theta)=E(\theta+2\pi)$, and it should also depends on $\theta$/N. Both of these properties can be achieved if we assume there is a level crossing in the ground state as $\theta$ is varied. At large N, the theory is thought to have a large number of meta-stable, Lorentz-invariant states. We label such states by $k$, such that the energy of each of these state is
\begin{equation}
E_{k}(\theta)=N^{2}g\left(\frac{\theta+2\pi k}{N}\right)\,.
\end{equation}
for some function $g$. The ground state energy is simply the minimal energy state,
\begin{equation}
E(\theta)=\min_{k}E_{k}\,.
\end{equation}
The function $E(\theta)$ is periodic, but not smooth. In particular, when $\theta=\pi$ there is a level crossing.
To understand the form of $E(\theta)$, we can use the fact that it has a minimum at $\theta=0$. This is due to fact that the Euclidean path integral is a sum over configurations weighted by $e^{i\theta\nu}$. Only for $\theta=0$ this is real and positive, hence maximizing $e^{-vE\left(\theta\right)}$, and so minimizing $E\left(\theta\right)$. By Taylor expansion, we expect 
\begin{equation}
E\left(\theta\right)=\min_{k}\frac{1}{2}C\left(\theta+2\pi k\right)^{2}+{\cal O}\left(\frac{1}{N}\right)
\end{equation}
where $C=\chi\left(0\right)/\left(16\pi^{2}N\right)^{2}$. 
Classically, CP is a good symmetry of the theory at $\theta=0,\pi$.
At $\theta=\pi$, there are two degenerate ground states and time-reversal
invariance maps one to the other. Therefore, at large N Yang-Mills,
time-reversal invariance is spontaneously broken at $\theta=\pi$. This coincides with the results for finite N using discrete anomalies as written below.
\\ \\
\\
\paragraph{Finite N argument from anomaly matching} In what follows, we review some of the theoretical arguments presented in~\cite{Gaiotto:2017yup}. We work in Euclidean signature, such that the coefficient of the $\theta$ parameter appears with an imaginary $i$ factor. In order to understand the phases of pure $SU\left(N\right)$ Yang-Mills
theory at $\theta=\pi$, it is useful to look at the role of the global
structure of the gauge group. We consider two different pure YM theories
with the following global group structure
\[
G=SU(N)\qquad\text{and}\qquad G=PSU(N)\equiv SU(N)/\mathbb{Z}_{N}\,.
\]
The theta angles take different ranges in these two cases:
\[
\theta_{SU\left(N\right)}\in\left[0,2\pi\right)\qquad\text{and}\qquad\theta_{SU\left(N\right)/\mathbb{Z}_{N}}\in\left[0,2\pi N\right)
\,.\]
Both theories are time reversal $\left({\cal T}\right)$ invariant 
at $\theta=0$ . Moreover, time reversal or parity cannot be spontaneously
broken at $\theta=0$ as shown in~\cite{Vafa:1984aa}. However, pure $SU(N)$
YM theory is also ${\cal T}$ invariant at $\theta=\pi$, while $SU(N)/\mathbb{Z}_{N}$
is ${\cal T}$ invariant at $\theta=\pi N$. Thus, gauging the $\mathbb{Z}_{N}$
centre of the $SU(N)$ gauge group, might break time reversal
invariance, which exactly fits to the concept of a mixed \textquoteright t
Hooft anomaly. The actual \textquoteright t Hooft anomaly at $\theta=\pi$
is between a one-form $\mathbb{Z}_{N}$ symmetry that we wish to gauge
and time reversal.

Starting with the usual $SU(N)$ YM theory at $\theta=\pi$, will
denote the $SU(N)$ gauge connection as $a$. Moreover, consider a gauge
connection of a $U(N)\cong\left(U(1)\times SU(N)\right)/\mathbb{Z}_{N}$
gauge group, denoted by ${\cal A}$, which decomposed into a $U(1)$ connection, $v$, and the $SU(N)$ gauge connection $a$:
\begin{equation}
{\cal A}=a+\frac{1}{N} v\,\mathbb{1}_{N\times N}\,,
\end{equation}
with a field strength
\begin{equation}
{\cal G}=d{\cal A}+{\cal A}\wedge{\cal A}\,.
\end{equation}
We couple this to a Background Field (BF) theory, which we write in the form
\begin{equation}
\L_{BF}=\frac{i}{2\pi}\int \ell\wedge\left(d v-NB\right)\,.
\end{equation}
$B$ is a two-form background field, while $\ell$ acts as a Lagrange multiplier. This BF theory exhibits a one-form gauge symmetry:
\begin{equation}
v\rightarrow v+N\lambda\qquad\text{and}\qquad B\rightarrow B+d\lambda\,.
\end{equation}
The field strength ${\cal G}$ is not invariant under the one-form
gauge symmetry of the BF theory, 
\begin{equation}
{\cal G}\rightarrow{\cal G}+d\lambda\,.
\end{equation}
Thus, the usual Yang-Mills term for ${\cal G}$ will not be gauge
invariant. Therefore, we must write a gauge invariant combination
${\cal G}-B$ as follows
\begin{align}
S_{SU\left(N\right)/Z_{N}} & =\frac{1}{2g^{2}}\int\text{Tr}\left({\cal G}-B\right)\wedge*\left({\cal G}-B\right)+\frac{i}{2\pi}\int \ell\wedge\left(d v-NB\right)\\
& \;+S_{\theta}\,.\nonumber 
\end{align}
The simplest gauge invariant theta term that one might wish to add to the theory is
\begin{equation}
S_{\theta}=\frac{i\theta}{8\pi^{2}}\int\text{Tr}\left({\cal G}-B\right)\wedge\left({\cal G}-B\right)
\end{equation}
for $\theta\in\left[0,2\pi\right)$. However, under the shift $\theta\rightarrow\theta+2\pi$
\begin{equation}
S_{\theta}\rightarrow S_{\theta}+\frac{i}{4\pi}\int\text{Tr}{\cal G}\wedge{\cal G}-\frac{i}{2\pi}\int\text{Tr}{\cal G}\wedge B+\frac{iN}{4\pi}\int\text{Tr}B\wedge B\,.
\end{equation}
The equation of motion for $\ell$ gives $\text{Tr}{\cal G}=d v=NB$.
Moreover, the first term in the above equation, $\frac{i}{4\pi}\int\text{Tr}{\cal G}\wedge{\cal G}$,
is related to the usual theta-term of the $SU(N)$ YM theory. It is
set by the third homotopy group of the gauge group which is $\pi_{3}\left(SU(N)\right)\cong\mathbb{Z}$.
Thus, the change in the action is 
\begin{equation}
\Delta S_{\theta}=-\frac{iN}{4\pi}\int\text{Tr}B\wedge B+i2\pi\mathbb{Z}\,.
\end{equation}
Therefore, the action is not invariant under $\theta\rightarrow\theta+2\pi$.
In order to fix this subtlety, we add a contact term for $B$, with a contact coupling $p$, which is also known as the discrete $\theta$ parameter,
\begin{equation}
S_{\theta}=\frac{i\theta}{8\pi^{2}}\int\text{Tr}\left({\cal G}-B\right)\wedge\left({\cal G}-B\right)-\frac{ipN}{4\pi}\int B\wedge B\,.\label{eq:Full_theta_term}
\end{equation}
As a result, the action is now invariant under 
\[
\theta\rightarrow\theta+2\pi\,,
\]
accompanied by the transformation
\[
p\rightarrow p-1\,.
\]
The contact term $\frac{ipN}{4\pi}\int B\wedge B$ is gauge invariant
if and only if $p\in\mathbb{Z}$, more exactly the discrete theta
angle $p$ can only take the values 
\[
p=0,1,...,N-1\,.
\]
This is exactly the result we expected from an $SU(N)/\mathbb{Z}_{N}$
YM theory as
\begin{equation}
\theta_{SU\left(N\right)/\mathbb{Z}_{N}}=\theta+2\pi p\;\in\left[0,2\pi N\right)\,.
\end{equation}
Therefore, if we start with $SU(N)$ YM theory at $\theta=\pi$ and
we gauge the one-form symmetry, we end up with an $SU(N)/\mathbb{Z}_{N}$
YM theory at
\begin{equation}
\theta_{SU\left(N\right)/\mathbb{Z}_{N}}=\pi\left(1+2p\right)\,.
\end{equation}
This theory is time reversal invariant only when $\theta_{SU\left(N\right)/\mathbb{Z}_{N}}=0,\pi N$. Thus, for an even $N$,  there is no choice of $p\in\mathbb{Z}$ such that $\pi\left(1+2p\right)=\pi N$. From this we conclude that if we gauge the one-form symmetry we lose time reversal invariance. In other words, there exists a mixed \textquoteright t Hooft anomaly between the $\mathbb{Z}_{N}$ one-form symmetry and time reversal. This \textquoteright t Hooft anomaly guarantees that the vacuum cannot be a trivial non-degenerate gapped state. \\
For odd $N$ the situation is different. One can choose a counter term, with $p=\frac{N-1}{2}$, such that time-reversal is not broken for both pure $SU(N)$ with $\theta = \pi$ and for $\theta_{SU\left(N\right)/\mathbb{Z}_{N}}$ with $SU(N)/\mathbb{Z}_{N} = \pi N$. However, one cannot choose simultaneously the same counter term such that time reversal is not broken for both $\theta=0$ and $\theta=\pi$. This is sometimes called in the litrature global inconsistency~\cite{Gaiotto:2017yup}. It follows that the theory at long distances cannot be trivial for all values of $\theta$.  The theory is believed to have a trivial confined ground state away from $\theta=\pi$~\cite{Karasik:2019bxn}. Thus, the consequences of this global inconsistency are in this particular case similar to the discussion above for even $N$. Therefore, with this assumption, time reversal symmetry should be spontaneously broken at $\theta=\pi$.

\bibliographystyle{JHEP}
\bibliography{theta_pi_AVIV}

\end{document}